\documentclass[aip,apl,reprint,showpacs,10pt,citeautoscript]{revtex4-2}
\usepackage{amsfonts}
\usepackage{hyperref}
\usepackage{amsmath,amssymb,epsfig,color}
\usepackage{float}
\usepackage{amsmath}
\usepackage{amssymb}
\usepackage{graphicx}%
\providecommand{\U}[1]{\protect\rule{.1in}{.1in}}
\AtBeginDocument{    \newwrite\bibnotes
\def\bibnotesext{Notes.bib}
\immediate\openout\bibnotes=\jobname\bibnotesext
\immediate\write\bibnotes{@CONTROL{REVTEX41Control}}
\immediate\write\bibnotes{@CONTROL{    apsrev41Control,author="08",editor="1",pages="1",title="0",year="1"}}
\if@filesw
\immediate\write\@auxout{\string\citation{apsrev41Control}}    \fi
}
\providecommand{\U}[1]{\protect\rule{.1in}{.1in}}
\providecommand{\U}[1]{\protect\rule{.1in}{.1in}}

\newcommand*\monbse[1]{{Mo$_{\rm x}$Nb$_{\rm 1 -x}$S$_2$ }}
\begin{document}
\title{Ising superconductivity: a first-principles perspective}
\author{Darshana Wickramaratne}
\affiliation{Center for Computational Materials Science, U.S. Naval Research Laboratory,
Washington, DC 20375, USA}
\author{I.I. Mazin}
\affiliation{Department of Physics and Astronomy, George Mason University, Fairfax, VA
22030, USA}
\affiliation{Quantum Science and Engineering Center, George Mason University, Fairfax, VA
22030, USA}
\date{\today }

\begin{abstract}
The recent discovery of Ising superconductivity has garnered a lot of interest due in part to the resilience of
these superconductors to large in-plane magnetic fields.  In this Perspective we explain the basic concepts that define the
behavior of Ising superconductors, provide an overview of the electronic structure and magnetic properties with a
focus on NbSe$_2$, summarize key experimental observations that have been made in this class of superconductors, highlight
the role that defects and proximity-induced effects at interfaces have on Ising superconductivity and finally discuss the 
prospects for observing Ising superconductivity in bulk materials.
\end{abstract}

\maketitle
\section{Introduction}
A fundamental concept in the theory of superconductors is the paramagnetic limiting magnetic field (Clogston-Chandrasekar-Pauli limiting field) \cite{clogston1962upper,chandrasekhar1962note}, 
henceforth referred to as the Pauli-limiting field.
This is the approximate magnetic field where superconductivity is suppressed 
when the Zeeman splitting of the spin degenerate states at the Fermi level
exceeds the magnitude of the superconducting gap.
Recently, superconductivity in a number of two-dimensional materials was shown to be 
surprisingly resilient to an in-plane magnetic field that greatly exceeded the Pauli limiting field.
The first report of this large in-plane critical field was in 
gated single monolayer MoS$_2$, a two-dimensional transition metal
dichalcogenide (TMD) \cite{lu2015evidence}.  Shortly thereafter large in-plane critical fields
that exceeded the Pauli limit were found in a number of other two-dimensional materials; including 
monolayer NbSe$_2$ \cite{xi2016Ising,sergio2018tuning}, TaS$_2$ \cite{sergio2018tuning},
and gated WS$_2$ \cite{lu2018full}.
Superconductors with thermodynamic critical fields, $H_{c2}$, that exceed the Pauli limiting field are often associated with
spin triplet superconductors (since the Cooper pairs with parallel spins having the ability to screen the applied
magnetic field), or due to extrinsic effects such as spin-orbit scattering.
The large in-plane critical magnetic fields in these two-dimensional materials, now referred to as Ising superconductors has a very different intrinsic origin.

The large in-plane $H_{c2}$ in the initial experiments on the 
monolayer TMDs was hypothesized to 
originate from the large spin-orbit splitting due to the transition metal in these TMDs.
This was based on a general understanding that the Fermi surface of these hexagonal materials
is comprised of two concentric Fermi rings around the 
K and K$^{\prime}$ points of the Brillouin zone.  The large SOC of the transition metal
splits the two
Fermi contours within a given K valley and the pseudospin direction is pinned along $\hat{z}$.
This description of the Fermi surface has formed the basis for a number of model theories
that have been put forth to analyze the experiments that have been reported
on Ising superconductors \cite{zhou2016ising,shaffer2020crystalline,mockli2018robust,sosenko2017unconventional}.
We note several authors have also provided an overview on the different materials where evidence for Ising
superconductivity has been uncovered \cite{zhang2021ising,wang2021ising,li2021recent}.
While these model theories have been used to both attempt to explain the wide range of experiments and also predict
new phenomena, they are unable to establish materials specific insight into different phenomena.

In this Perspective we offer a personal viewpoint on how first-principles calculations can and have been
used to elucidate the properties of Ising superconductors.
Our aim is not to provide a comprehensive overview of the field of Ising superconductivity given that this remains
an active area of research.
Rather, our goal is to show how first-principles calculations
enable insight into materials-specific issues that are otherwise unattainable with
model Hamiltonian descriptions of the electronic structure of the TMD monolayers.
While Ising superconductivity has been reported in a range of materials,
our discussion will primarily focus on the example of the monolayer NbSe$_2$, the most widely studied Ising
superconductor.

Our discussion will highlight the hierarchy of energy scales that provides a natural explanation for the large
$H_{c2}$ in these materials, 
the presence of magnetic fluctuations \cite{isingprx,divilov2021magnetic}, which we show are present in the TMDs that exhibit
Ising superconductivity, and a detailed description of the fermiology including the role of the third pocket at the zone-center.
 No material is ideal, including the TMD Ising superconductors.  We will show that
non-magnetic and magnetic point defects which are likely to be present in these materials can manifest themselves in interesting phenomena such
as apparent disorder-induced enhancement in superconductivity \cite{zhao2019disorder,rubio2020visualization},
hysteresis in the superconducting phase \cite{idzuchi2021unconventional},
oscillations in the superconducting current and apparent breaking of rotational symmetry \cite{cho2022nodal,hamill2021two}.

This article is organized as follows - in Section \ref{sec:general} we provide a general discussion of the physics of
Ising superconductors, \ref{sec:electronic} we provide an overview of the electronic and magnetic
properties of monolayer NbSe$_2$, in Section \ref{sec:defects} we discuss the impact that defects, doping and alloying have on the superconducting
properties of Ising superconductors, in Section \ref{sec:heterostructures} we show how experiments involving heterostructures with Ising
superconductors lead to a number of puzzling results, and in Section \ref{sec:bulk} we show how Ising superconductivity a phenomenon often associated with single monolayers
can be found in bulk materials.

\section{General considerations}
\label{sec:general}
We first discuss at a general level the physics that leads to Ising superconductivity.
We start from the simplest description of the Fermi surface of monolayer TMDs where
Ising superconductivity is observed - two 
quasi-circular Fermi contours around the K and K$^{\prime}$ valleys which are related by
time reversal symmetry.
\begin{figure*}[!t]
\includegraphics[width=14.5cm]{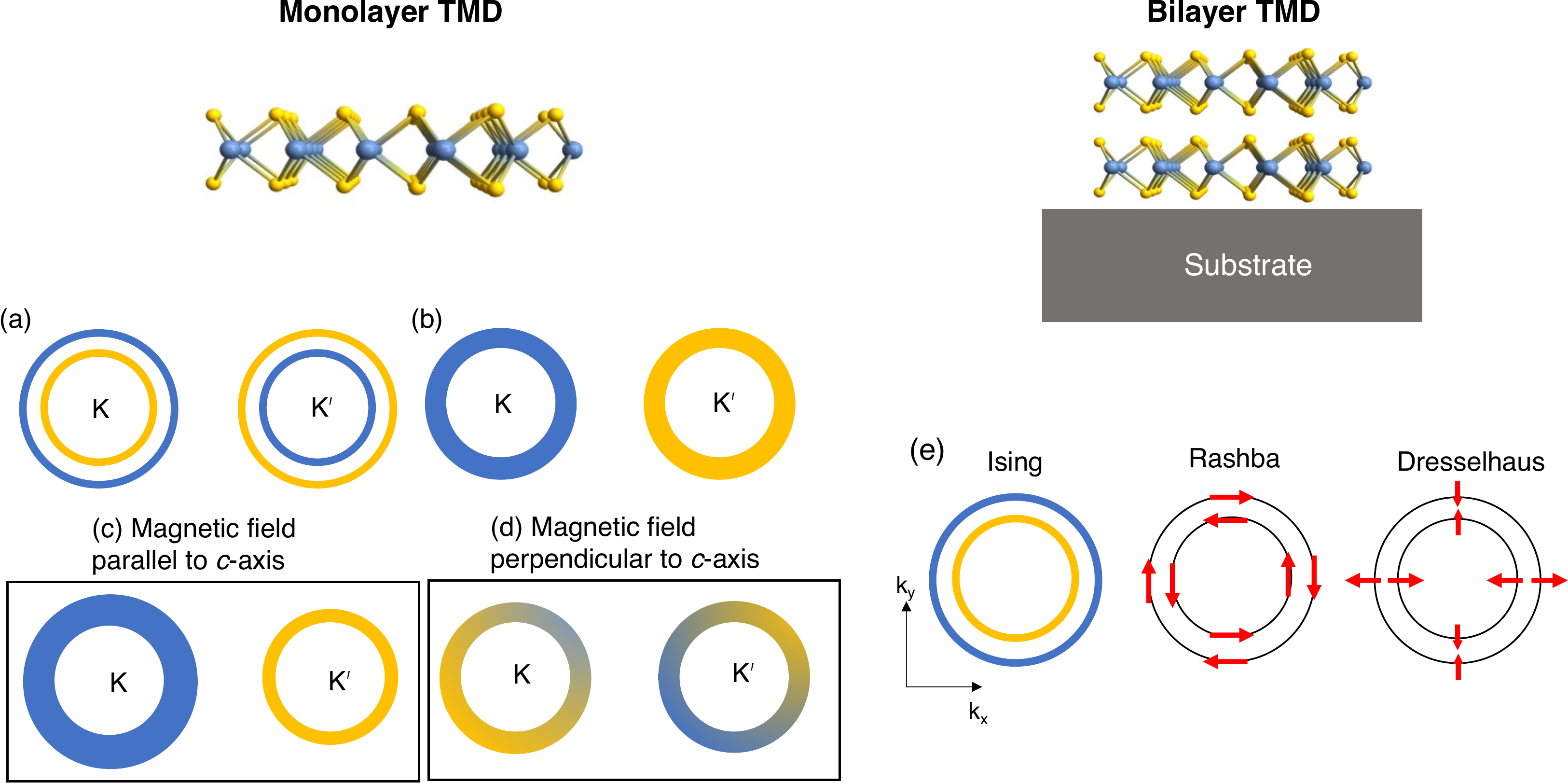}
\caption{(a) Schematic illustration of the Fermi contours around K and K$^{\prime}$ in a monolayer TMD Ising superconductor in the presence
of spin-orbit interaction.  Blue represents $\hat{z}$ pseudospin states and yellow represents $-\hat{z}$ pseudospin states. 
(b) Net pseudospin enclosed around each Fermi contour denoted by the area enclosed by the blue and yellow shaded regions. 
(c)  Fermi contours and net pseudospin in the presence of a magnetic
field that is parallel to the $c$-axis of the monolayer, i.e parallel to the direction of the spins.  (d) Fermi contours 
and the net pseudospin in the presence
of an in-plane magnetic field, i.e field is perpendicular to the direction of the spins.  Note that the area bounded by the two Fermi contours within a given
valley is similar in (b) and (d) but different in (c) due to the Zeeman splitting.
(e) Possible in-plane spin textures (Rashba or Dresselhaus) in addition to the Ising spin texture in monolayer or few-layer TMDs due to symmetry
lowering effects, eg. substrate-induced effects as illustrated in the top right.}
\label{fig:fermi}%
\end{figure*}

With spin-orbit interaction the spin-degenerate states within a given valley are split by a spin-orbit coupling (SOC)
parameter, $\Delta_{\rm SOC}$, resulting in two concentric Fermi contours as illustrated
in Figure \ref{fig:fermi}(a).
The pseudospin along $\hat{z}$ flips direction between the two Fermi contours such that within a given valley
there is a finite polarization of pseudospins as denoted by the colored area in Figure \ref{fig:fermi}(b).  Moving from one valley (K)
to its time-reversal partner (K$^{\prime}$) the direction of the pseudospin flips as illustrated by the different colored areas 
in Figure \ref{fig:fermi}(b) so that the overall
net magnetic moment is zero.

If we apply a magnetic field, $H$, parallel to the $c$-axis of the monolayer TMD, the Fermi contours within a given
valley are Zeeman split by an amount that is linearly proportional to the magnitude of the magnetic field.  Assuming the
magnetic field is applied along $+\hat{z}$, the Fermi contours derived from pseudospin $+\hat{z}$ will increase in area
while those derived from $-\hat{z}$ will decrease.  Hence, based on our notation in Figure \ref{fig:fermi}(a), the area
between the two concentric Fermi contours will now be different between K and K$^{\prime}$ as illustrated in Figure \ref{fig:fermi}(c).
This manifests itself in a finite net spin polarization as expected by an amount that is proportional to the magnetic field and the single-spin density
of states of the material (neglecting the Stoner enhancement).

When the magnetic field is perpendicular to the $c$-axis, the spins that are $\alt\Delta_{\rm SOC}$ away from the Fermi level
tilt away from $\hat{z}$.  
Since $\Delta_{\rm SOC}$ is significantly larger than the magnitude of the Zeeman splitting 
and is also larger than the superconducting gap, $\Delta$, the tilting of the spins away from $\hat{z}$ means the spin-susceptibility
is defined by states that are approximately an energy $\Delta_{\rm SOC}$ away from the Fermi level.  To first order in the magnitude of the
in-plane magnetic field there is no Zeeman splitting of the Fermi contours.  As a result the area between the two Fermi contours within a single valley
does not change.  However, there is a net spin polarization due to the fact that the 
pseudospins are tilted away from $\hat{z}$ to be in plane in the two valleys.  

With this heuristic understanding of the response of the electronic structure to out-of-plane and in-plane
magnetic fields it is now easy to address why Ising superconductors exhibit such a large anisotropy in their $H_{c2}$.   
The $H_{c2}$ of a superconductor is determined by the difference
in the free energies of the normal state and the superconducting state of a material.  Within BCS theory and at 0 K
the thermodynamic critial field, $H_{c0}$ is defined as
$F_{n}-F_{s}\sim\Delta^{2}N(0)/2=(\chi_{n}-\chi_{s})H_{0}^{2}/2$
where $N(0$) is the density of states at the Fermi level.
Since the spin susceptibility for magnetic fields parallel to the $c$-axis is determined
by the Zeeman splitting of the Fermi contours, superconductivity is suppressed once the magnitude of the Zeeman splitting
exceeds $\Delta$, $i.e$, the superconductivity is Pauli limited.
In contrast, when the magnetic field is perpendicular to the $c$-axis the spin susceptibility is determined by states
approximately $\Delta_{\rm SOC}$ from the Fermi level.  Since $\Delta_{\rm SOC}$ is significantly larger than the magnitude of
the Zeeman splitting induced by the applied magnetic field, $H$, and $\Delta$, the
$H_{c2}$ is formally infinite in an Ising superconductor.

From this discussion it is apparent that the hierarchy of energy scales that dictates
this large in-plane $H_{c2}$ is $H \leq \Delta \ll \Delta_{\rm SOC}$. 
The combination of time-reversal symmetry and broken inversion symmetry guarantees that
the pseudospin has pure $\pm \hat{z}$ character around the K valleys. 
Moving from a monolayer
to a bilayer leads to four bands that cross
the Fermi level, forming two pairs degenerate in energy and  comprised of equal contributions from both
monolayers.  Within a single K valley the sign of the pseudospin flips from $+\hat{z}$ to $-\hat{z}$ between 
the pair of degenerate states.  The second pair of degenerate states is an energy $\Delta_{\rm SOC}$ away.
This layer degeneracy of the states within a K valley is due to
the centrosymmetric stacking of the two monolayers.  The vertical stacking between the two monolayers
leads to a finite interlayer hopping, $t$, between the monolayers.  We will show in Section \ref{sec:electronic} that this
interlayer hopping is $k$-dependent.  

Additional terms can arise that impact the energy and spin character of these states.
Indeed Shaffer {\it et al.} \cite{shaffer2020crystalline} have shown that the coupling between a single monolayer
and a substrate can introduce an in-plane component to the spin texture (Rashba and/or Dresselhaus) as illustrated in Figure \ref{fig:fermi}(e)
in addition to the
Ising component along $\hat{z}$. 
These considerations also apply to the bilayer structures.  One additional consideration is changes
in stacking away from the ground state centrosymmetric stacking, which can lead to changes in the
interlayer hopping, as well as to a non-centrosymmetric bilayer structure. 
Since the ground state stacking of the bulk NbSe$_2$ unit cell is centrosymmetric,
there is often an implicit assumption that  Ising superconductivity cannot survive in bilayer
or bulk structures.
However in Section \ref{sec:bulk} we highlight experimental studies where 
evidence of Ising superconductivity in bulk compounds has indeed been observed and point to work
that indicates there is a region of the parameter space that involves
$\Delta_{\rm SOC}$, interlayer hopping and the coupling to a substrate/superstrate that leads to Ising protection
in bilayer structures.  

These general considerations provide qualitative insight into two hallmarks of Ising superconductivity, the large in-plane $H_{c2}$ that greatly
exceeds the Pauli-limiting field and the reduction in $H_{c2}$ for few-layer and bulk films.  In the following sections we will discuss materials-specific
aspects of the electronic properties that provide insight that cannot be attained using the considerations we have detailed above.

\section{Electronic structure}
\label{sec:electronic}
For the discussion that follows we will focus on monolayer NbSe$_2$, which is the most widely studied Ising superconductor.
\begin{figure*}[!t]
\includegraphics[width=12.5cm]{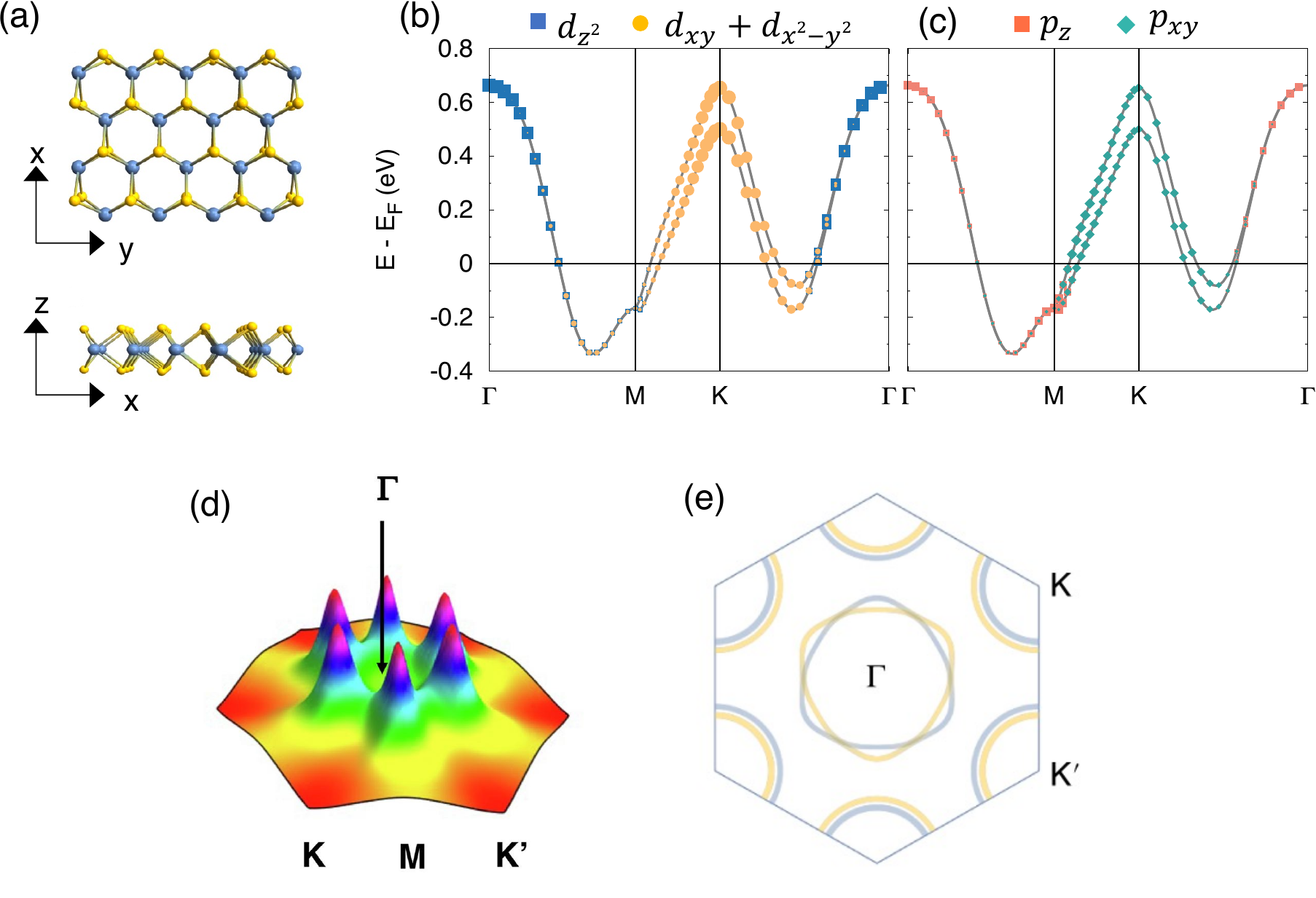}
\caption{Electronic properties of monolayer NbSe$_2$.  (a)  Top view and side view of monolayer NbSe$_2$ where Nb atoms (blue)
are in a trigonal prismatic coordination with the Se atoms (yellow).  (b)  Band structure with SOC calculated with DFT
of monolayer NbSe$_2$.  The colors correspond to the contribution by the different $d$ orbitals denoted in the legend above.
(c) Renormalized spin-susceptbility as a function of spin spiral {\bf q} vector across
the Brillouin zone of monolayer NbSe$_2$.  
(d) The
Fermi contour of NbSe$_2$ where the colors denote the different pseudospin character of the states.
Panel (d) was obtained from Ref.~\onlinecite{das2021renormalized}}
\label{fig:bandstructure}%
\end{figure*}
The building block for monolayer NbSe$_2$ is a Nb atom that is in a trigonal
prismatic coordination with the Se atoms.  The overall structure is non-centrosymmetric and belongs to space group
P$\bar{6}$m2.  The trigonal crystal field splits the $4d$ states of Nb$^{4+}$ into the following
groups, $d_{z^{2}}$, [$d_{x^{2}-y^{2}}$, $d_{xy}$] and [$d_{xz}$,$d_{yz}$], which in the absence of SOC
leads to a spin degenerate band that crosses the Fermi level
several times.  
This band structure, which has been reported in several studies
leads to three Fermi contours, one that encircles $\Gamma$ due to contributions from the Nb $d_{z^{2}}$ states 
and a pair of Fermi contours that encircle that K and K$^{\prime}$ points
of the Brillouin zone that are contributed by the Nb [$d_{x^{2}-y^{2}}$, $d_{xy}$] states.
The combination of SOC due to Nb and the lack of an inversion center leads to
momentum-dependent spin-orbit splitting everywhere except along the $\Gamma$-M line.
The electronic structure of monolayer NbSe$_2$ is summarized in Figure \ref{fig:bandstructure}.
Note that there is also a minor admixture of the Se $p$-states, where at $\Gamma$ the Se $p_z$
states contribute while at K the Se $p_{x,y}$ states contribute as illustrated in Figure \ref{fig:bandstructure}(b).

To understand this momentum dependent spin-orbit splitting that pins the pseudospins to be along $\hat{z}$
we consider that the bands that cross the Fermi level a state at a given momentum $k$ can be defined as
$\left\vert {\phi}\right\rangle =\eta\left\vert d_{x^{2}-y^{2}}\right\rangle
+\beta\left\vert d_{xy}\right\rangle +\gamma\left\vert d_{z^{2}}\right\rangle$
where $\eta^{2}+\beta^{2}+\gamma^{2}=1$. Here, we ignore the minor contribution of the chalcogen $p$-states
to the bands that cross the Fermi level.  Note that $d_{z^{2}}$ corresponds to
$\left\vert l,m\right\rangle =\left\vert 2,0\right\rangle ,$ $d_{x^{2}-y^{2}}$
to $(\left\vert 2,2\right\rangle +\left\vert 2,-2\right\rangle )\sqrt{2}$, and
$d_{xy}$ to $(\left\vert 2,2\right\rangle -\left\vert 2,-2\right\rangle
)/\mathrm{i}\sqrt{2}$, where $l$ is the angular momentum quantum number and $m$ is the
magnetic quantum number.

When we account for spin, the Hamiltonian at each $k$-point is a 
($2\times2$) matrix.  Since the single monolayers have $z/-z$ mirror symmetry around
the Nb atom, the Hamiltonian does not
include contributions from the $\left\vert 2,\pm1\right\rangle $ orbitals.
Hence, the nondiagonal matrix elements $L_{\pm}$ are zero. However, 
the diagonal element can be defined as $L_{z}=2(\eta\operatorname{Im}\beta
-\beta\operatorname{Im}\eta).$ One phase can always be selected as real, for
instance, $\eta,$ which leads to  $L_{z}=2\eta\operatorname{Im}\beta.$ 
Only along $\Gamma$-M is the value of $\beta$ real by symmetry which leads to $L_{z}=0$ and zero splitting due to SOC.  Elsewhere along
the Brillouin zone,
this leads to an orbital moment that can only be parallel or antiparallel with respect to 
$\hat{z}$.  With the inclusion of spin-orbit coupling this leads to the pseudospin to 
point along $\hat{z}$, which is indeed what we find in electronic structure calculations of monolayer NbSe$_2$.
The spin-orbit induced splitting,
which is maximal at K (and K$^{\prime}$), depends largely on the transition metal element of the TMD monolayer.
From DFT calculations, the splitting at K is 150 meV in monolayer NbSe$_2$, in monolayer NbS$_2$ it is 115 meV, while it is 280 meV
in monolayer TaS$_2$.

One of the experimental challenges when studying single monolayer films is that conventional bulk probes
of the electronic structure of materials are not readily applied to single monolayer films.
However, one can still obtain valuable information on the properties of the single monolayer by comparing the calculated
and experimentally measured properties of the bulk compounds in addition to comparisons of
trends associated with calculated properties of the single monolayer versus bulk.
Hence, it is instructive to briefly discuss the properties
of bulk NbSe$_2$ --- a well known superconductor although not an Ising superconductor.

The bulk NbSe$_2$ unit cell is comprised of two monolayers of NbSe$_2$ that are vertically stacked leading to a centrosymmetric structure that belongs
to space group $P6_3/mmc$.  
Superconductivity which has been measured below 7.2 K in bulk NbSe$_2$ \cite{frindt1972superconductivity} is widely thought to
be driven entirely by electron-phonon coupling \cite{boaknin2003heat,noat2015quasiparticle}.
First-principles calculations of the Fermi surface of bulk NbSe$_2$ at the level of
the generalized gradient approximation are generally in agreement
with ARPES measurements \cite{johannes2006fermi,rossnagel2001fermi}. 

One experimental fact that had largely been overlooked are spin susceptibility
measurements of bulk NbSe$_2$, which reports a low temperature spin 
susceptibility of 3.04$\times$10$^{-4}$ emu/mole \cite{iavarone2008effect}.
The Pauli susceptibility of bulk NbSe$_2$ using the DOS at the
Fermi level from DFT calculations 
yields a spin susceptibility that is a factor of 3 lower than experiment.
This is an indication of a considerable Stoner renormalization, corresponding to a 
Stoner factor $I\approx \frac{2}{3}N(0)$. On the other hand, the calculated Stoner-renormalized
susceptibility is about 40\% {\em larger} than the experimental number\cite{isingprx}.

This overestimation  of the calculated spin susceptibility in comparison to experiment
is a well known consequence of using a mean-field theory such as the generalized gradient approximation
implementation of DFT to calculate the properties of itinerant metals that are close a magnetic
instability.  Other examples of this overestimation in the calculated spin response at the DFT level
in comparison to experiment include studies on bulk Pd \cite{larson2004magnetism} and the 
iron-based superconductors \cite{mazin2008problems}.
The magnetic moments and the tendency towards long-range magnetic order, 
which are overestimated at this mean field level in itinerant metals, are in reality suppressed
due to the presence of long range magnetic fluctuations \cite{moriya2012spin}.

This difference in the calculated Pauli susceptibility and the experimental susceptibility of bulk NbSe$_2$
therefore points to the presence of strong
spin fluctuations in NbSe$_2$, which would renormalize the Pauli paramagnetic susceptibility.
Further indirect evidence for the presence of spin fluctuations in the bulk structure is that
state-of-the-art 
first-principles calculations of the superconducting T$_c$ that only account for electron-phonon coupling largely overestimate
$T_c$ compared to the experimentally established value \cite{das2022electron}.  This discrepancy is likely due to pair-breaking effects
such as spin fluctuations not being accounted for in the calculations. 

Returning to the monolayer structure, the calculated ferromagnetic spin susceptibility of monolayer NbSe$_2$ has been calculated to be a factor of 1.5
larger than the bulk - suggesting spin fluctuations are, not unexpectedly,
stronger in the single monolayer compared to bulk NbSe$_2$ \cite{isingprx}.
Das {\it et al.} \cite{das2021renormalized} have also convincingly shown that monolayer NbSe$_2$ exhibits
antiferromagnetic spin fluctuations with a {\bf q} vector of [0.2,0,0] as illustrated in Figure \ref{fig:bandstructure}(d).
The role of magnetism in NbSe$_2$ has also been highlighted by a number of
theoretical studies by other groups \cite{divilov2021magnetic,costa2022ising}.

\begin{figure}[!t]
\includegraphics[width=8.5cm]{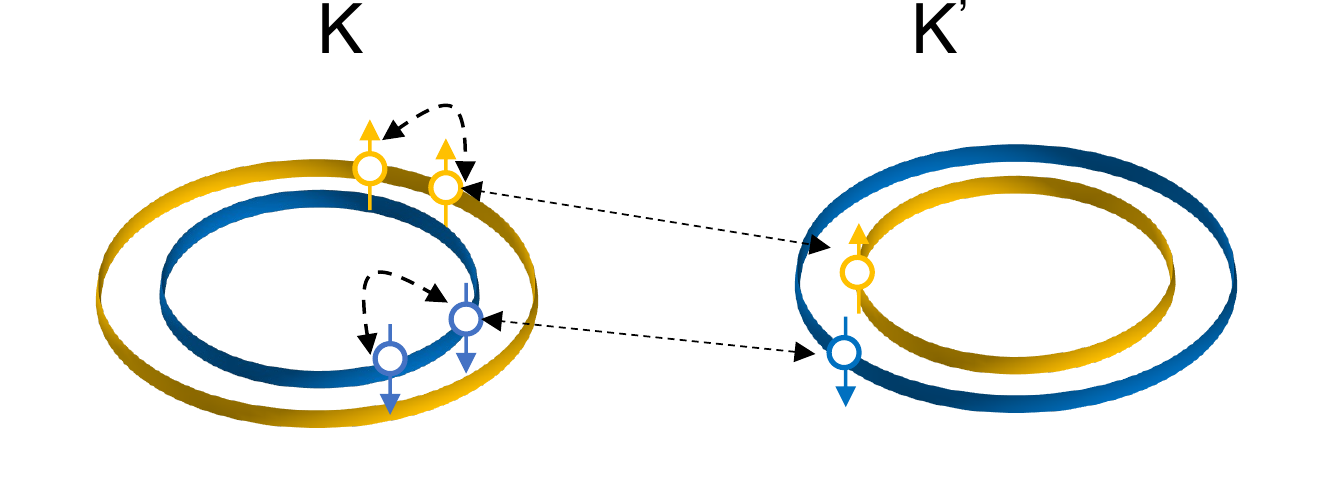}
\caption{Schematic illustration of Fermi contours around {\rm K} and K${\prime}$ illustrating the possible pairing interactions involving phonons.  The yellow
and blue contours represent states with pseudospin $m_z=1+$ or $m_z=1-$, respectively.  The black dotted arrows denote intravalley and intervalley pairing interactions
that can occur due to electron-phonon interaction.}
\label{fig:pairing}%
\end{figure}

There are several ramifications of this combination of the electronic structure and the presence of spin fluctuations on the
 Ising superconductivity.
Arguably the most interesting ramifications of the presence of these spin fluctuations are on the pairing interactions.  
If we only consider Cooper pairs formed from states that reside at K and K$^{\prime}$ (i.e neglecting contributions at $\Gamma$), 
in the simplest approximation, 
one may assume that the amplitude of the order parameter across the two Fermi contours
 illustrated in Figure \ref{fig:pairing} is similar.  
Since the pseudospins flip sign between K
and K$^{\prime}$, the formation of Cooper pairs due to phonons can only 
involve processes that do not require a 
change in the sign of the pseudospin.
Superconducting states are either classified as singlet to triplet depending on whether the total
spin of the Cooper pair is 0 or 1.
One interesting consequence of the broken Kramer's degeneracy at the K and K${\prime}$ valleys in the single monolayer TMDs is that the
superconducting state is neither single nor triplet but a combination of singlet {\it and} triplet \cite{isingprx}.

To put these qualitative arguments on more firm theoretical footing recent first-principles calculations \cite{das2022electron}
have determined the impact of electron-phonon interaction and spin fluctuations on the pairing interactions
in monolayer NbSe$_2$.  The principal findings of this study is that (1) the electron-phonon interaction is highly 
anisotropic and the dominant pairing mechanism is due to 
intervalley processes between the Fermi contours with similar pseudospins at the
K and K${\prime}$ valleys (solid black arrows in Figure \ref{fig:pairing}), (2) the $T_c$ calculated due to electron-phonon
interactions alone is greatly overestimated compared to the experimental $T_c$, and (3) spin fluctuations weaken the strength of pairing
interactions and bring the calculated $T_c$ in closer agreement with experiment.
The fact that the calculated ferromagnetic spin susceptibility of the monolayer structure is larger than bulk NbSe$_2$ may
also explain why the $T_c$ of the bulk structure is larger than that of the monolayer.  

Other competing mechanisms have also been proposed to be at play and impact superconductivity.  
The role of the charge density wave (CDW) phase in NbSe$_2$,
and its impact on superconductivity
continues to be actively debated.  Some theoretical studies have proposed that the CDW phase is responsible
for the reduction in the superconducting $T_c$ compared to the $T_c$ obtained entirely due to electron-phonon
interaction \cite{zheng2019electron}.  However, recent experiments have shown that the suppression of the CDW
in monolayer NbSe$_2$ due to alloying with Mo also leads to a suppression of the superconducting $T_c$ \cite{wan2022observation},
which is incompatible with the proposal in Ref.~\onlinecite{zheng2019electron}. 
Furthermore applying pressure \cite{leroux2015strong} and introducing defects by irradiation \cite{cho2018using} have been shown to suppress the CDW while 
leading to a minor change in the superconducting transition temperature.

Recent tunneling measurements on monolayer NbSe$_2$, which observed a number of satellite peaks on either side of the
primary coherence peaks, were interpreted as a manifestation of a Leggett mode between a singlet $s$-wave
and a spin-triplet $f$-wave channel \cite{wan2022observation}.
We note however, that this postulation is incompatible with the findings of first-principles calculations \cite{das2022electron}
where it was shown that the $f$-wave pairing interactions within the K valley are weak.  
Other tunneling measurements in the presence of an in-plane magnetic
field have interpreted their measurements by pointing to the possible presence of a subleading spin-triplet order parameter \cite{kuzmanovic2022tunneling}.
Transport measurements in the presence of an in-plane
magnetic field found a surprising two-fold periodicity in the magnetoresistance which was interepreted as evidence of a 
competing nematic superconducting instability \cite{hamill2021two} that coexists with a conventional singlet superconducting state.
In order to confirm the experimental manifestation of these proposed mechanisms another issue to contend with is the role of
defects and disorder, which we discuss in the following section.

\section{Defects, doping and alloying}
\label{sec:defects}
\begin{figure*}[!t]
\includegraphics[width=12.5cm]{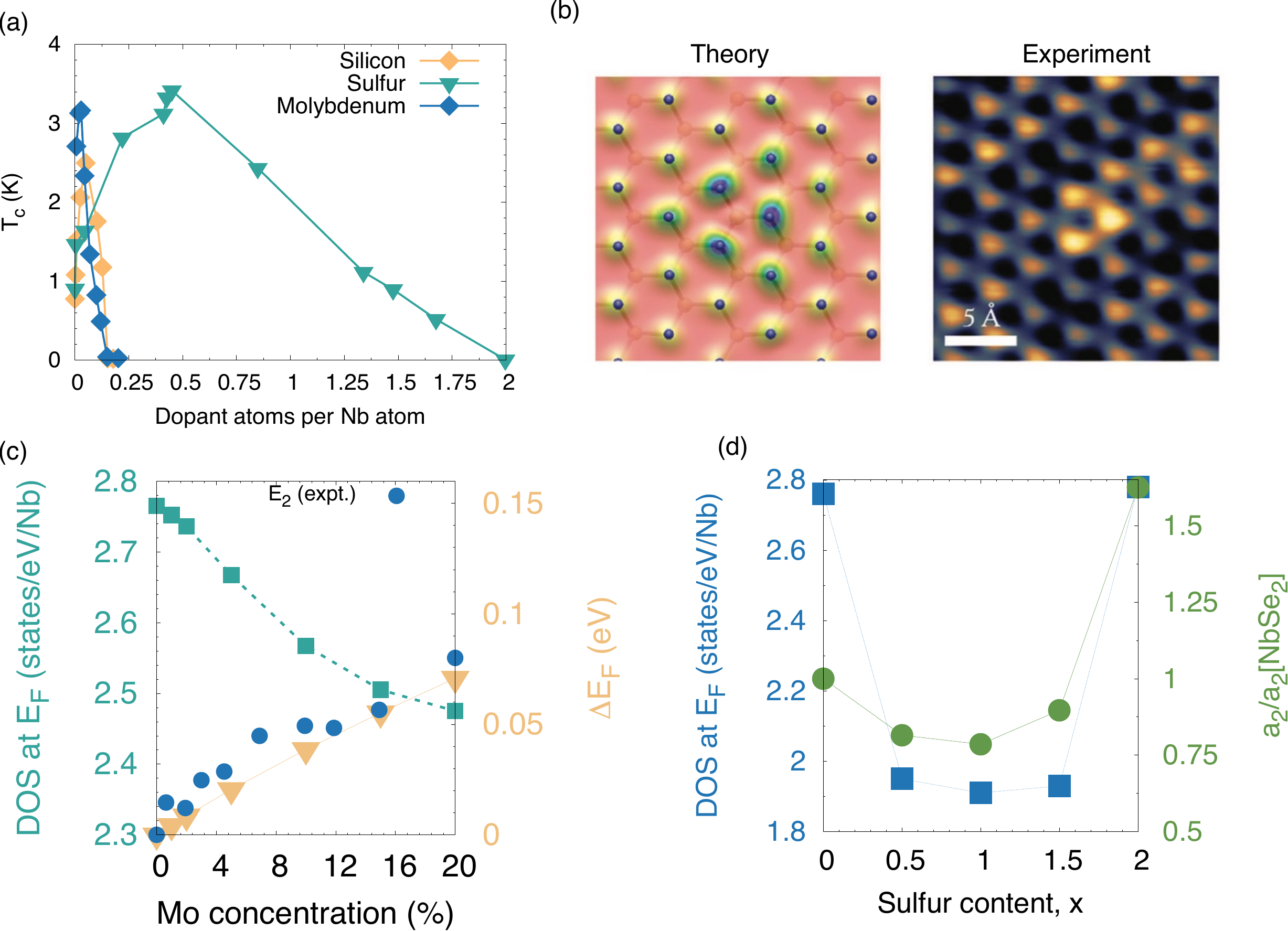}
\caption{Effect of alloying and doping in monolayer NbSe$_2$.  (a)  Summary of experimental reports on change in $T_c$ as a function of alloying monolayer NbSe$_2$
with sulfur, silicon and molybdenum. (b)  Comparison of STM images from experiment and DFT calculations for substitutional Mo in NbSe$_2$.  The isosurface
can be related to the .  (c)  First-principles calculations of the effect of alloying Mo in NbSe$_2$ on the density of states at the Fermi level and the shift in the Fermi
level as a function of Mo content.  Data points from experiment are included as blue dots.  (d)  First-principles calculations of the effect of alloying on the Se site with
S on the density of states (left vertical axis) and the magnitude of ferromagnetic spin fluctutations normalized with respect to monolayer NbSe$_2$.
Panel (b-d) was obtained from Refs.~\onlinecite{nbse2_fractal_natcomm,wan2022nontrivial}.}
\label{fig:doping}%
\end{figure*}

Experiments on the role of disorder either through alloying, doping or defects have led to a number of puzzling results.  
Alloying NbSe$_2$ with sulfur (which is isovalent to selenium) was found to change $T_c$ non-monotonically
with sulfur content --- a pronounced increase in $T_c$ up to a critical sulfur content subsequently followed by a monotonic suppression. 
Qualitatively similar non-monotonic changes in $T_c$ was found in NbSe$_2$ that had been exposed to silicon \cite{zhao2019disorder}. 
This increase in $T_c$ for intermediate concentrations
of the alloying element was postulated to be evidence of fractal superconductivity --- $i.e.,$ a disorder-induced enhancement of superconductivity.
However, this purported enhancement was reported with respect to NbSe$_2$ samples that had a $T_c$ of $\sim$1 K --- significantly lower than the widely
reported value of 3-4 K.
Finally, doping NbSe$_2$ with Mo was found to enhance $T_c$ slightly up to a critical concentration of doping
after which superconductivity was suppressed \cite{wan2022nontrivial}.  These changes in $T_c$ with doping in monolayer NbSe$_2$ is summarized
in Figure \ref{fig:doping}(a).

Interpreting the experiments on alloying or doping requires information on where in the lattice the impurity is incorporated 
since this determines the electrical properties
of the impurity. 
For example when Mo is incorporated in NbSe$_2$ it incorporates substitutionally on the Nb site as shown by STM studies \cite{wan2022nontrivial}
and further corroborated by first-principles calculations (cf. Figure \ref{fig:doping}(b)).
The extra electron from Mo dopes NbSe$_2$.  
This leads to a narrow range of doping where $T_c$ is initially enhanced beyond which $T_c$ gradually
decreases and superconductivity is suppressed at a critical Mo composition. 
Doping NbSe$_2$ with Mo leads to a monotonic reduction in the density of states which leads to two 
effects.  It decreases the magnitude of the electron-phonon coupling that is responsible for pairing which suppresses
$T_c$ and it would also suppress the magnitude of pair-breaking magnetic fluctuations which would enhance $T_c$ as illustrated
in Figure \ref{fig:doping}(c).
It is likely that a combination of these two changes as a function of doping leads to the non-monotonic
change in $T_c$ that has been observed in experiments where NbSe$_2$ is doped with Mo.

A number of experiments have also explored the impact of alloying NbSe$_2$ with isovalent elements across the entire composition range showing
a surprising enhancement in $T_c$ as illustrated in Figure \ref{fig:doping}(a).
At this point it is useful to recall that based on Anderson's theorem, non-magnetic impurities will not lead to pair-breaking in a conventional 
$s$-wave superconductor.  However, magnetic impurities are pair breaking.  
The tendency towards magnetism in these materials suggests native point defects may lead to finite
magnetic moments.  Indeed, first-principles calculations have shown selenium vacancies to have low formation energies \cite{nbse2_fractal_natcomm}.  These calculations have also uncovered
a large modulation of the spin density that extends several lattice sites away from the vacancy with a magnetic
moment amplitude of $\sim$ 0.5 $\mu_B$ \cite{nbse2_proximity}.  
If selenium vacancies are magnetic point defects in NbSe$_2$, this would provide
a natural explanation for a number of experimental puzzles observed when monolayer NbSe$_2$ has been alloyed with sulfur and silicon.

First, selenium vacancies can act as a source of scattering and lower the
residual resistivity ratio.  The magnetic nature of these vacancies would also render them pair-breaking leading to a lower $T_c$. 
This also coincides with the fact that in the experiments low values of the residual resistivity ratio
were found in samples where the $T_c$ of monolayer NbSe$_2$ was low \cite{cho2018using}.
Furthermore, in the experiments where sulfur and silicon were alloyed into NbSe$_2$ it was assumed that the NbSe$_2$ monolayer
prior to alloying was stoichiometric and in the case of silicon that Si was being adsorbed on the surface \cite{zhao2019disorder}.
Our first-principles calculations have shown that this assumption is incorrect and depending on the concentration of selenium vacancies
that are incorporated during growth, sulfur and silicon can occupy these vacant selenium sites \cite{nbse2_fractal_natcomm}.  This can lead
to non-monotonic changes in the electronic and magnetic properties of NbSe$_2$ as a function of alloy content as shown in Figure \ref{fig:doping}(d) and it
is likely these non-monotonic changes in the electronic and magnetic properties that lead to non-monotonic change in $T_c$, not fractal superconductivity.

\begin{figure}[!ht]
\includegraphics[width=8.5cm]{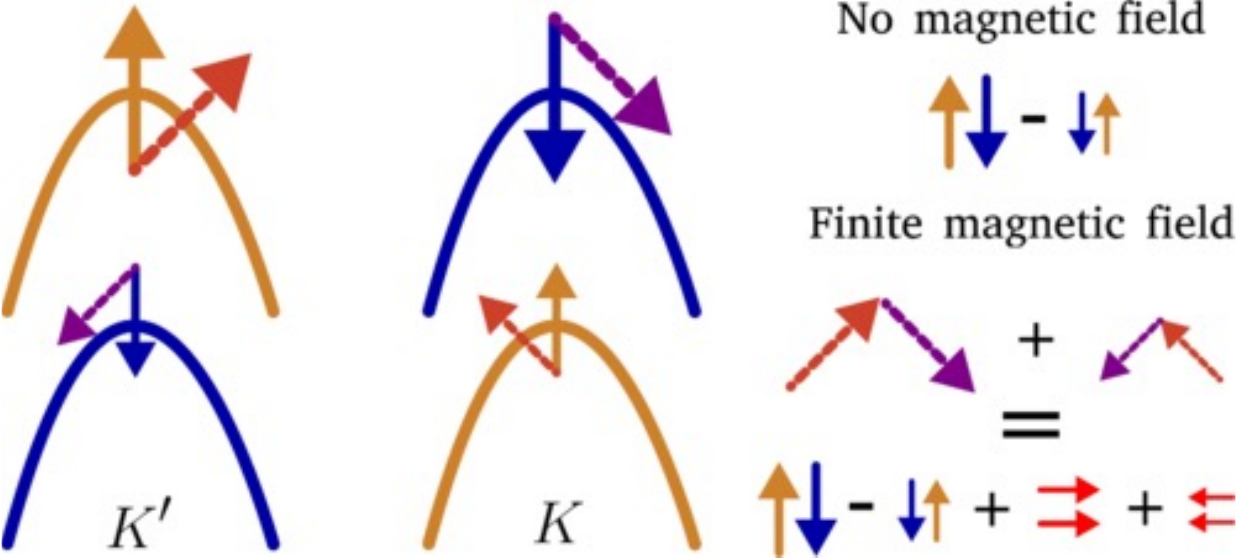}
\caption{Schematic illustration of the effect of a magnetic field on the canting of the spins at the 
K and K$^{\prime}$ points.
Figure obtained from Ref.~\onlinecite{mockli2020Ising}.}
\label{fig:scattering}%
\end{figure}

Non-magnetic defects in the prescence of a magnetic field can also have nontrivial effects in Ising superconductors.
Let us first consider an Ising superconductor with non magnetic impurities in the absence of a magnetic field.  
Intervalley scattering between the two outer (or inner) contours
is not permitted since this requires a spin flip.  If $H$ is parallel to $\hat{z}$, there is Zeeman splitting which suppresses superconductivity
but the conditions that prohibit intervalley scattering between the contours where a spin flip is required remains the same.  Hence, no scattering
occurs since the pseudospin remains pinned to $\hat{z}$.  When $H$  is in-plane
the spins around the K and K$^{\prime}$ valleys tilt in plane in the direction of $H$ 
as discussed in Section \ref{sec:general} and illustrated in Figure \ref{fig:scattering}, acquiring a triplet component.  

Since the pseudospins now have a finite in-plane component
intervalley scattering between the outer (and inner) contours due to non-magnetic impurities 
is allowed due to the finite overlap of spin states.  Since the degree by which the 
pseudospins tilt away from $\hat{z}$ to be in-plane is proportional to the magnitude of the applied in-plane magnetic field, the magnitude of the scattering is expected to increase
as a function of the magnitude of the magnetic field.  This interplay between the Ising spin-orbit coupling, non magnetic impurities and scattering
induced by an in-plane magnetic field would be expected to lead to a broadening of the coherence peaks associated with tunneling due to an Ising
superconductor.
Indeed this prediction manifests itself in recent tunneling experiments that have been performed on NbSe$_2$
tunneling heterostructures where CrBr$_3$ was used as a magnetic insulator tunnel barrier, which we discuss in the following section.

\section{Heterostructures with magnetic and non-magnetic insulators}
\label{sec:heterostructures}
\begin{figure}[!ht]
\includegraphics[width=8.5cm]{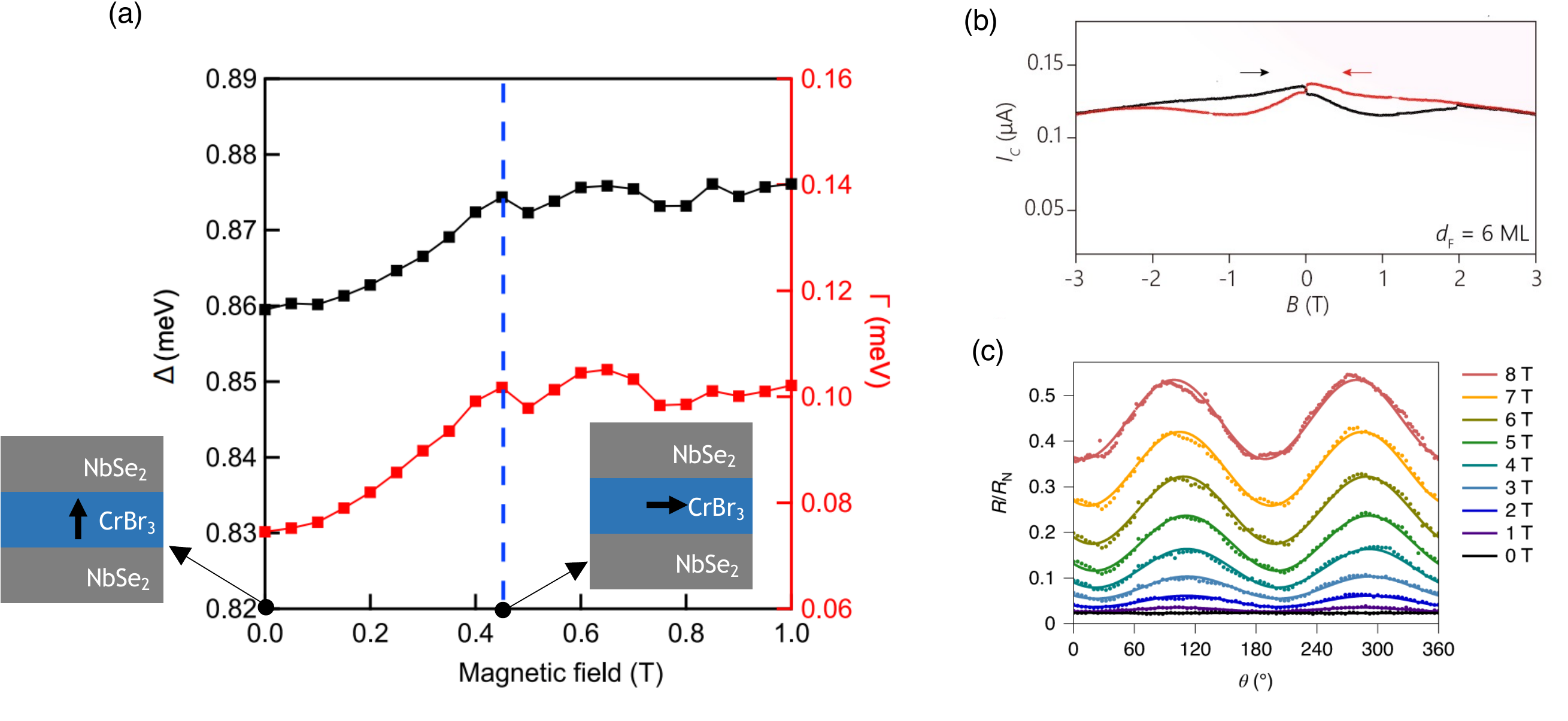}
\caption{Tunneling measurements in NbSe$_2$-magnetic insulator heterostructures.  (a) Schematic illustration of the device geometry used
in tunneling experiments along with measurements of the change in the superconducting gap and coherence peaks as a function of in-plane
magnetic field reported in Ref.~\onlinecite{kang2021giant}, (b) Hysteresis in the switching current of the NbSe$_2$/Cr$_2$Ge$_2$Te$_6$/NbSe$_2$
junction as a function of applied in-plane magnetic field, and (c) Field dependence of the magnetoresistance in a 
NbSe$_2$/CrBr$_3$ heterostructure.
Figures adapted from Ref.~\onlinecite{kang2021giant,hamill2021two,idzuchi2021unconventional}.
}
\label{fig:tunneling}%
\end{figure}

Since the Ising superconductors are two-dimensional materials they also represent an ideal platform to explore
the role of interfaces and the interaction of Ising superconductivity with different phenomenon such as magnetism,
charge-density waves, and topological order.  
One such heterostructure are atomically thin Josephson junction heterostructures where the Ising superconductor
is used as the top and bottom superconducting contacts sandwiched between an insulating barrier such as WSe$_2$ \cite{kuzmanovic2022tunneling}
or magnetic insulating barriers such as 
CrBr$_3$ \cite{hamill2021two}, to Cr$_2$Ge$_2$Te$_6$ \cite{idzuchi2021unconventional,kang2022van,ai2021van} as summarized in Figure \ref{fig:tunneling}. 
In the case of the experiments that use WSe$_2$ as a tunnel barrier, the tunneling measurements that were performed
in the presence of an in-plane magnetic field find evidence of equal spin-singlet-triplet pairs in the tunneling conductance.

In the case of the experiments with the magnetic insulating barriers, even through the barrier materials are different there are a 
number of unifying observations.  This includes the observation of an apparent pair-breaking of $\Delta$ but a surprising strengthening of the coherent peak
width in tunneling, hysterersis that sets in below the superconducting $T_c$, evidence of broken rotational symmetry that occurs
only in the superconducting state, and spin-filtering tunneling processes \cite{idzuchi2021unconventional,kang2022van,ai2021van,hamill2021two}
One idea put forth to explain the apparent symmetry breaking is the possibility of a two-component order parameter where
a nematic phase coexists with the Ising superconducting phase in NbSe$_2$.  It was postulated that the formation of such a heterostructure
can couple to both order parameters and the nematic phase would be responsible for the rotational symmetry 
breaking observed in the experiments \cite{cho2022nodal}.
Another possible explanation for this rotational symmety breaking is the role played by nonmagnetic defects which in the presence of an in-plane magnetic
field can behave as pair-breaking defects and lead to signatures that are consistent with those reported in experiment \cite{nbse2_proximity,mockli2020Ising}.

An alternative approach to explore magnetic-Ising superconductor interfaces is to replace the magnetic
layer that is in proximity with the Ising superconductor with magnetic transition metal ions that are 
intercalated.  Such approaches have been explored experimentally in few-layer NbSe$_2$ and NbS$_2$ layers
where magnetic ions such as Cr and Fe are intercalated in between the layers \cite{wu2022highly,haley2021long,xie2022structure}.  The intercalated
ions have been shown to form an ordered superlattice within the basal plane of the TMD.  These superlattices
such as Fe$_{1/3}$NbS$_{2}$ exhibit degenerate magnetic states \cite{wu2022highly}
that are tunable by the concentration of
magnetic ions that are intercalated. 
The impact of these intercalated ions and magnetic phases on superconductivity remains to be addressed.  
One intriguing possibility is the potential to observe
signatures of Ising superconductivity in these intercalated structures given that Ising superconductivity can occur
in bulk strucures as we will discuss in the next section. 

\section{Ising superconductivity in bulk materials}
\label{sec:bulk}
The principal signature of Ising superconductivity is a large in-plane $H_{c2}$ which is pronounced
in single monolayers of the superconducting TMDs.  
Surprisingly, experiments have shown a range of bulk materials that are comprised of the TMDs exhibit large values of in-plane $H_{c2}$
 that are comparable to those found in single monolayers.  This includes experiments performed
on misfit compounds that contain monolayers of NbS$_2$ and NbSe$_2$ sandwiched in between LaS and LaSe layers
\cite{kashihara1979upper,coleman1983dimensional,samuely2021extreme,leriche2021misfit,devarakonda2021signatures}, 
twisted monolayers of TaS$_2$ \cite{ma2018unusual},
bulk TaS$_2$ intercalated with different organic molecules \cite{prober1980upper,gamble1970superconductivity} and 
the 4H, 3R, and 6R polymorphs of TaSe$_2$ and TaS$_2$ \cite{xing2021extrinsic,ribak2020chiral,achari2022alternating,tanaka2020superconducting} .
The early reports of these large in-plane $H_{c2}$ values was interepreted as arising from spin-orbit
scattering (using short spin-orbit scattering times) or the presence of a Rashba-like spin texture of the states at the Fermi surface which would enable in principle
finite-momentum pairing \cite{devarakonda2021signatures}.
A majority of these proposed interpretations were made without taking into account the electronic structure of these misfit compounds
and assessing whether the details of the Fermi surface would in fact favor Ising superconductivity.

\begin{figure}[!ht]
\includegraphics[width=8.5cm]{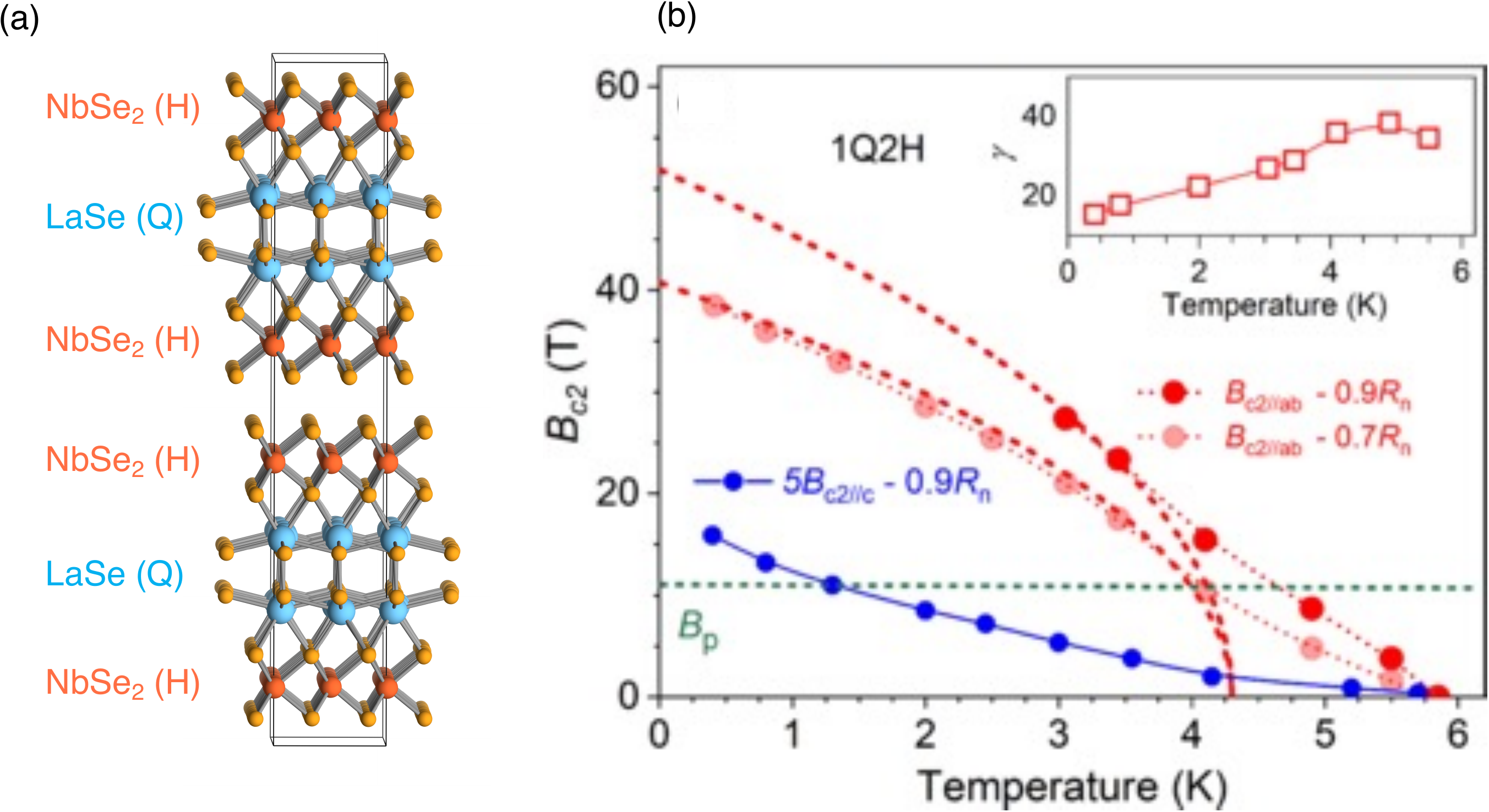}
\caption{Ising superconductivity in bulk compounds.  (a)  Schematic illustration of the 
(LaSe)$_{\rm 1.14}$(NbSe$_2$)$_2$ misfit compound where evidence of large in-plane critical fields that exceed
the Pauli limit were meaured.  (b)  Experimental reports of the in-plane critical field as a function of
temperature for bulk misfit compounds that contain NbSe$_2$ layers.  Figure (b) was obtained from Reference \onlinecite{samuely2021extreme}.}
\label{fig:misfit}%
\end{figure}

We suggest that in each of these experiments the large in-plane $H_{c2}$ is likely a manifestation of Ising
superconductivity in bulk compounds.  
The ground state stacking of the metallic TMDs is 2Ha, which is centrosymmetric with a center of inversion between the layers.
In some misfit compounds, the misfit layer is intercalated between the pair of TMD monolayers thereby weakening the interlayer
coupling of the bilayer structure.  In other misfit compounds such as (LaSe)$_{\rm 1.14}$(NbSe$_2$)$_2$, which is schematically
illustrated in Figure \ref{fig:misfit}(a), the misfit layer is intercalated between pairs of the bilayer TMD that are stacking in their equilibrium
2H$_a$ stacking.  In both configurations the misfit layer leads to doping of the TMD layers.  First-principles calculations combined
with photoemission data have recently shown that the states at the Fermi level in these misfit compounds is best described as a heavily 
electron-doped TMD layer \cite{leriche2021misfit}.  
Interestingly these bulk misfit compounds exhibit in-plane $H_{c2}$ values that greatly exceed the Pauli limit \cite{samuely2021extreme}.

A natural question then is what leads to the large values of $H_{c2}$ in these bulk compounds.  
In the case of the misfit compounds where the misfit layer is intercalated in between the bilayer TMD, the increase
in the interlayer separation is sufficient to weaken the interlayer coupling so that the TMDs in the misfit mimic act as 
two decoupled monolayers.  Even though these monolayers are doped, the Fermi level shifts to an energy where the Ising protection
is still present.
In the misfit compounds where the misfit layer is in between a pair of bilayers one aspect to consider is that the misfit layer
above and below the bilayer is stacked asymmetrically.  This can lead to an
asymmetric on-site potential that effectively breaks the layer degeneracy
that is otherwise present in the electronic structure of the bilayer.  
Identifying the competition between the relevant energy scales --- $\Delta_{\rm SOC}$, interlayer coupling between the TMD layers with different
stacking configurations of the misfit layer --- and the combined effect of all of this on the details of the Fermi surface are parameters
that can be extracted using first-principles calculations.  Indeed there is a range of parameter
space where indeed bilayer and bulk misfit compounds comprised of the TMDs can exhibit signatures of Ising superconductivity such as large
in-plane $H_{c2}$ \cite{samuely2023protection}.
Given the wide range of misfit compounds where large values of in-plane $H_{c2}$ 
have been observed \cite{kashihara1979upper,coleman1983dimensional,samuely2021extreme,leriche2021misfit,devarakonda2021signatures,ng2022misfit,klemm2015pristine}, 
we expect these calculations when analyzed in the context of the experiments 
will play a key role in unraveling this signature of Ising superconductivity in these bulk compounds.

\begin{figure}[!ht]
\includegraphics[width=8.5cm]{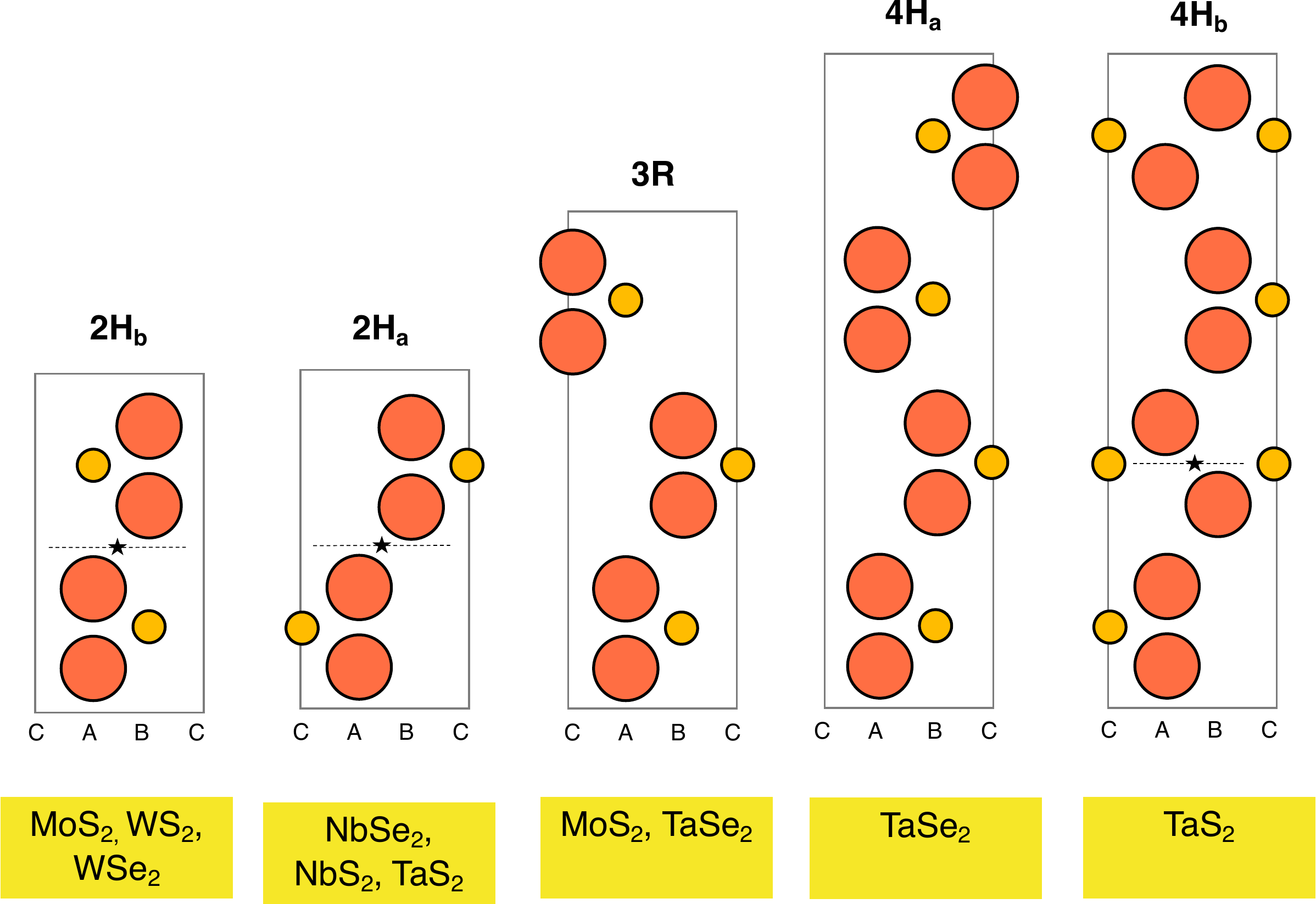}
\caption{Different polytypes of the bulk transition metal dichalcogenides.  The transition metal is denoted
with the small yellow circles and the chalcogens with the large red circles.  The labels A, B, and C at the bottom
can be used to determine the stacking sequence of each polytype.  For the centrosymmetric stacking configurations, the
center of inversion is denoted with a black star and a black dotted line.  The materials where these polytypes have been
identified are listed in a yellow box.  Figure adapted from Ref.\onlinecite{brown1965layer}.
}
\label{fig:polytype}%
\end{figure}

Another route to access Ising superconductivity in bulk materials is considering the fact that
while the ground state stacking of the bulk TMDs is centrosymmetric there are a large number
of polytypes that these materials can exist in due to the different stacking configurations that are possible.
We list some of these polytypes in Table \ref{tab:polytype} and schematically illustrate their stacking configurations
in Figure \ref{fig:polytype}.  
\begin{table}[]
\begin{tabular}{l|cccc}
Polytype & \begin{tabular}[c]{@{}l@{}}Space group \\ (\#)\end{tabular} & Centrosymmetric? & Symmorphic?  \\
\hline \hline
2H       & \begin{tabular}[c]{@{}l@{}}P6$_{3}$/mmc   \\ (190)\end{tabular}  & $\checkmark$     & $\times$     \\ [0.15ex]
3R       & \begin{tabular}[c]{@{}l@{}}R3m   \\ (160)\end{tabular}      & $\times$         & $\checkmark$     \\ [0.15ex]
4Ha      & \begin{tabular}[c]{@{}l@{}}P3m1  \\  (156)\end{tabular}     & $\times$         & $\checkmark$ \\ [0.15ex]
4Hb      & \begin{tabular}[c]{@{}l@{}}P6$_{3}$/mmc  \\  (194)\end{tabular}  & $\checkmark$     & $\times$    \\ [0.15ex]
\hline \hline
\end{tabular}
\caption{List of some of the possible polytypes of the bulk TMDs, their space group, whether they are centrosymmetric or not
and if they belong to a symmorphic space group or not.}
\label{tab:polytype}
\end{table}
We note that some of these polytypes such as the 3R and 4Ha configuration are not centrosymmetric
and indeed there have been experimental reports of Ising superconductivity in these bulk 
polytypes \cite{klemm2015pristine}.

\section{Conclusions}
Since the first report of Ising superconductivity in two-dimensional materials in 2015 the number of materials that exhibit large in-plane 
magnetic fields which is a signature of Ising superconductivity has grown to include a wide range of low-dimensional materials. 
There is sometimes a myth (which we show to be unfounded) that Ising superconductivity is a rather esoteric phenomenon.
In this Perspective article we discuss how Ising superconductivity can in fact be understood at the basic level using our general understanding
of the electronic structure of Ising superconductor materials and thermodynamic arguments for the susceptibility in the normal and superconducting state.
Beyond providing this heuristic understanding, one of the major themes of this Perspective article is that first-principles calculations provide 
materials-specific microscopic insight into the wide variety of experiments that have been perfomed thus far on Ising superconductors.
We highlighted the important role played by non-magnetic and magnetic point defects and proximity induced effects at interfaces in
Ising superconductor heterostructures.
Finally we show that Ising superconductivity is not a phenomenon that only occurs in two-dimensional materials - there are a wide range of bulk materials
where Ising superconductivity is likely to occur as well.
\acknowledgements
D.W was supported by the Office of Naval Research through the Naval Research Laboratory's Basic Research Program.
I.I.M was supported by the Office of Naval Research through grant N00014-20-1-2345.


{\bf References}


\begin{thebibliography}{63}%
\makeatletter
\providecommand \@ifxundefined [1]{%
 \@ifx{#1\undefined}
}%
\providecommand \@ifnum [1]{%
 \ifnum #1\expandafter \@firstoftwo
 \else \expandafter \@secondoftwo
 \fi
}%
\providecommand \@ifx [1]{%
 \ifx #1\expandafter \@firstoftwo
 \else \expandafter \@secondoftwo
 \fi
}%
\providecommand \natexlab [1]{#1}%
\providecommand \enquote  [1]{``#1''}%
\providecommand \bibnamefont  [1]{#1}%
\providecommand \bibfnamefont [1]{#1}%
\providecommand \citenamefont [1]{#1}%
\providecommand \href@noop [0]{\@secondoftwo}%
\providecommand \href [0]{\begingroup \@sanitize@url \@href}%
\providecommand \@href[1]{\@@startlink{#1}\@@href}%
\providecommand \@@href[1]{\endgroup#1\@@endlink}%
\providecommand \@sanitize@url [0]{\catcode `\\12\catcode `\$12\catcode
  `\&12\catcode `\#12\catcode `\^12\catcode `\_12\catcode `\%12\relax}%
\providecommand \@@startlink[1]{}%
\providecommand \@@endlink[0]{}%
\providecommand \url  [0]{\begingroup\@sanitize@url \@url }%
\providecommand \@url [1]{\endgroup\@href {#1}{\urlprefix }}%
\providecommand \urlprefix  [0]{URL }%
\providecommand \Eprint [0]{\href }%
\providecommand \doibase [0]{https://doi.org/}%
\providecommand \selectlanguage [0]{\@gobble}%
\providecommand \bibinfo  [0]{\@secondoftwo}%
\providecommand \bibfield  [0]{\@secondoftwo}%
\providecommand \translation [1]{[#1]}%
\providecommand \BibitemOpen [0]{}%
\providecommand \bibitemStop [0]{}%
\providecommand \bibitemNoStop [0]{.\EOS\space}%
\providecommand \EOS [0]{\spacefactor3000\relax}%
\providecommand \BibitemShut  [1]{\csname bibitem#1\endcsname}%
\let\auto@bib@innerbib\@empty
\bibitem [{\citenamefont {Clogston}(1962)}]{clogston1962upper}%
  \BibitemOpen
  \bibfield  {author} {\bibinfo {author} {\bibfnamefont {A.~M.}\ \bibnamefont
  {Clogston}},\ }\href@noop {} {\bibfield  {journal} {\bibinfo  {journal}
  {Phys. Rev. Lett.}\ }\textbf {\bibinfo {volume} {9}},\ \bibinfo {pages} {266}
  (\bibinfo {year} {1962})}\BibitemShut {NoStop}%
\bibitem [{\citenamefont {Chandrasekhar}(1962)}]{chandrasekhar1962note}%
  \BibitemOpen
  \bibfield  {author} {\bibinfo {author} {\bibfnamefont {B.}~\bibnamefont
  {Chandrasekhar}},\ }\href@noop {} {\bibfield  {journal} {\bibinfo  {journal}
  {Appl. Phys. Lett.}\ }\textbf {\bibinfo {volume} {1}},\ \bibinfo {pages} {7}
  (\bibinfo {year} {1962})}\BibitemShut {NoStop}%
\bibitem [{\citenamefont {Lu}\ \emph {et~al.}(2015)\citenamefont {Lu},
  \citenamefont {Zheliuk}, \citenamefont {Leermakers}, \citenamefont {Yuan},
  \citenamefont {Zeitler}, \citenamefont {Law},\ and\ \citenamefont
  {Ye}}]{lu2015evidence}%
  \BibitemOpen
  \bibfield  {author} {\bibinfo {author} {\bibfnamefont {J.}~\bibnamefont
  {Lu}}, \bibinfo {author} {\bibfnamefont {O.}~\bibnamefont {Zheliuk}},
  \bibinfo {author} {\bibfnamefont {I.}~\bibnamefont {Leermakers}}, \bibinfo
  {author} {\bibfnamefont {N.~F.}\ \bibnamefont {Yuan}}, \bibinfo {author}
  {\bibfnamefont {U.}~\bibnamefont {Zeitler}}, \bibinfo {author} {\bibfnamefont
  {K.~T.}\ \bibnamefont {Law}},\ and\ \bibinfo {author} {\bibfnamefont
  {J.}~\bibnamefont {Ye}},\ }\href@noop {} {\bibfield  {journal} {\bibinfo
  {journal} {Science}\ }\textbf {\bibinfo {volume} {350}},\ \bibinfo {pages}
  {1353} (\bibinfo {year} {2015})}\BibitemShut {NoStop}%
\bibitem [{\citenamefont {Xi}\ \emph {et~al.}(2016)\citenamefont {Xi},
  \citenamefont {Wang}, \citenamefont {Zhao}, \citenamefont {Park},
  \citenamefont {Law}, \citenamefont {Berger}, \citenamefont {Forr{\'o}},
  \citenamefont {Shan},\ and\ \citenamefont {Mak}}]{xi2016Ising}%
  \BibitemOpen
  \bibfield  {author} {\bibinfo {author} {\bibfnamefont {X.}~\bibnamefont
  {Xi}}, \bibinfo {author} {\bibfnamefont {Z.}~\bibnamefont {Wang}}, \bibinfo
  {author} {\bibfnamefont {W.}~\bibnamefont {Zhao}}, \bibinfo {author}
  {\bibfnamefont {J.-H.}\ \bibnamefont {Park}}, \bibinfo {author}
  {\bibfnamefont {K.~T.}\ \bibnamefont {Law}}, \bibinfo {author} {\bibfnamefont
  {H.}~\bibnamefont {Berger}}, \bibinfo {author} {\bibfnamefont
  {L.}~\bibnamefont {Forr{\'o}}}, \bibinfo {author} {\bibfnamefont
  {J.}~\bibnamefont {Shan}},\ and\ \bibinfo {author} {\bibfnamefont {K.~F.}\
  \bibnamefont {Mak}},\ }\href@noop {} {\bibfield  {journal} {\bibinfo
  {journal} {Nat. Phys.}\ }\textbf {\bibinfo {volume} {12}},\ \bibinfo {pages}
  {139} (\bibinfo {year} {2016})}\BibitemShut {NoStop}%
\bibitem [{\citenamefont {Sergio}\ \emph {et~al.}(2018)\citenamefont {Sergio},
  \citenamefont {Sinko}, \citenamefont {Gopalan}, \citenamefont {Sivadas},
  \citenamefont {Seyler}, \citenamefont {Watanabe}, \citenamefont {Taniguchi},
  \citenamefont {Tsen}, \citenamefont {Xu}, \citenamefont {Xiao},\ and\
  \citenamefont {Hunt}}]{sergio2018tuning}%
  \BibitemOpen
  \bibfield  {author} {\bibinfo {author} {\bibfnamefont {C.}~\bibnamefont
  {Sergio}}, \bibinfo {author} {\bibfnamefont {M.~R.}\ \bibnamefont {Sinko}},
  \bibinfo {author} {\bibfnamefont {D.~P.}\ \bibnamefont {Gopalan}}, \bibinfo
  {author} {\bibfnamefont {N.}~\bibnamefont {Sivadas}}, \bibinfo {author}
  {\bibfnamefont {K.~L.}\ \bibnamefont {Seyler}}, \bibinfo {author}
  {\bibfnamefont {K.}~\bibnamefont {Watanabe}}, \bibinfo {author}
  {\bibfnamefont {T.}~\bibnamefont {Taniguchi}}, \bibinfo {author}
  {\bibfnamefont {A.~W.}\ \bibnamefont {Tsen}}, \bibinfo {author}
  {\bibfnamefont {X.}~\bibnamefont {Xu}}, \bibinfo {author} {\bibfnamefont
  {D.}~\bibnamefont {Xiao}},\ and\ \bibinfo {author} {\bibfnamefont
  {B.}~\bibnamefont {Hunt}},\ }\href@noop {} {\bibfield  {journal} {\bibinfo
  {journal} {Nat. Comm.}\ }\textbf {\bibinfo {volume} {9}},\ \bibinfo {pages}
  {1427} (\bibinfo {year} {2018})}\BibitemShut {NoStop}%
\bibitem [{\citenamefont {Lu}\ \emph {et~al.}(2018)\citenamefont {Lu},
  \citenamefont {Zheliuk}, \citenamefont {Chen}, \citenamefont {Leermakers},
  \citenamefont {Hussey}, \citenamefont {Zeitler},\ and\ \citenamefont
  {Ye}}]{lu2018full}%
  \BibitemOpen
  \bibfield  {author} {\bibinfo {author} {\bibfnamefont {J.}~\bibnamefont
  {Lu}}, \bibinfo {author} {\bibfnamefont {O.}~\bibnamefont {Zheliuk}},
  \bibinfo {author} {\bibfnamefont {Q.}~\bibnamefont {Chen}}, \bibinfo {author}
  {\bibfnamefont {I.}~\bibnamefont {Leermakers}}, \bibinfo {author}
  {\bibfnamefont {N.~E.}\ \bibnamefont {Hussey}}, \bibinfo {author}
  {\bibfnamefont {U.}~\bibnamefont {Zeitler}},\ and\ \bibinfo {author}
  {\bibfnamefont {J.}~\bibnamefont {Ye}},\ }\href@noop {} {\bibfield  {journal}
  {\bibinfo  {journal} {Proceedings of the National Academy of Sciences}\
  }\textbf {\bibinfo {volume} {115}},\ \bibinfo {pages} {3551} (\bibinfo {year}
  {2018})}\BibitemShut {NoStop}%
\bibitem [{\citenamefont {Zhou}\ \emph {et~al.}(2016)\citenamefont {Zhou},
  \citenamefont {Yuan}, \citenamefont {Jiang},\ and\ \citenamefont
  {Law}}]{zhou2016ising}%
  \BibitemOpen
  \bibfield  {author} {\bibinfo {author} {\bibfnamefont {B.~T.}\ \bibnamefont
  {Zhou}}, \bibinfo {author} {\bibfnamefont {N.~F.}\ \bibnamefont {Yuan}},
  \bibinfo {author} {\bibfnamefont {H.-L.}\ \bibnamefont {Jiang}},\ and\
  \bibinfo {author} {\bibfnamefont {K.~T.}\ \bibnamefont {Law}},\ }\href@noop
  {} {\bibfield  {journal} {\bibinfo  {journal} {Phys. Rev. B}\ }\textbf
  {\bibinfo {volume} {93}},\ \bibinfo {pages} {180501} (\bibinfo {year}
  {2016})}\BibitemShut {NoStop}%
\bibitem [{\citenamefont {Shaffer}\ \emph {et~al.}(2020)\citenamefont
  {Shaffer}, \citenamefont {Kang}, \citenamefont {Burnell},\ and\ \citenamefont
  {Fernandes}}]{shaffer2020crystalline}%
  \BibitemOpen
  \bibfield  {author} {\bibinfo {author} {\bibfnamefont {D.}~\bibnamefont
  {Shaffer}}, \bibinfo {author} {\bibfnamefont {J.}~\bibnamefont {Kang}},
  \bibinfo {author} {\bibfnamefont {F.}~\bibnamefont {Burnell}},\ and\ \bibinfo
  {author} {\bibfnamefont {R.~M.}\ \bibnamefont {Fernandes}},\ }\href@noop {}
  {\bibfield  {journal} {\bibinfo  {journal} {Phys. Rev. B}\ }\textbf {\bibinfo
  {volume} {101}},\ \bibinfo {pages} {224503} (\bibinfo {year}
  {2020})}\BibitemShut {NoStop}%
\bibitem [{\citenamefont {M{\"o}ckli}\ and\ \citenamefont
  {Khodas}(2018)}]{mockli2018robust}%
  \BibitemOpen
  \bibfield  {author} {\bibinfo {author} {\bibfnamefont {D.}~\bibnamefont
  {M{\"o}ckli}}\ and\ \bibinfo {author} {\bibfnamefont {M.}~\bibnamefont
  {Khodas}},\ }\href@noop {} {\bibfield  {journal} {\bibinfo  {journal} {Phys.
  Rev. B}\ }\textbf {\bibinfo {volume} {98}},\ \bibinfo {pages} {144518}
  (\bibinfo {year} {2018})}\BibitemShut {NoStop}%
\bibitem [{\citenamefont {Sosenko}, \citenamefont {Zhang},\ and\ \citenamefont
  {Aji}(2017)}]{sosenko2017unconventional}%
  \BibitemOpen
  \bibfield  {author} {\bibinfo {author} {\bibfnamefont {E.}~\bibnamefont
  {Sosenko}}, \bibinfo {author} {\bibfnamefont {J.}~\bibnamefont {Zhang}},\
  and\ \bibinfo {author} {\bibfnamefont {V.}~\bibnamefont {Aji}},\ }\href@noop
  {} {\bibfield  {journal} {\bibinfo  {journal} {Phys. Rev. B}\ }\textbf
  {\bibinfo {volume} {95}},\ \bibinfo {pages} {144508} (\bibinfo {year}
  {2017})}\BibitemShut {NoStop}%
\bibitem [{\citenamefont {Zhang}\ and\ \citenamefont
  {Falson}(2021)}]{zhang2021ising}%
  \BibitemOpen
  \bibfield  {author} {\bibinfo {author} {\bibfnamefont {D.}~\bibnamefont
  {Zhang}}\ and\ \bibinfo {author} {\bibfnamefont {J.}~\bibnamefont {Falson}},\
  }\href@noop {} {\bibfield  {journal} {\bibinfo  {journal} {Nanotechnology}\
  }\textbf {\bibinfo {volume} {32}},\ \bibinfo {pages} {502003} (\bibinfo
  {year} {2021})}\BibitemShut {NoStop}%
\bibitem [{\citenamefont {Wang}, \citenamefont {Xu},\ and\ \citenamefont
  {Duan}(2021)}]{wang2021ising}%
  \BibitemOpen
  \bibfield  {author} {\bibinfo {author} {\bibfnamefont {C.}~\bibnamefont
  {Wang}}, \bibinfo {author} {\bibfnamefont {Y.}~\bibnamefont {Xu}},\ and\
  \bibinfo {author} {\bibfnamefont {W.}~\bibnamefont {Duan}},\ }\href@noop {}
  {\bibfield  {journal} {\bibinfo  {journal} {Acc. Mater. Res.}\ }\textbf
  {\bibinfo {volume} {2}},\ \bibinfo {pages} {526} (\bibinfo {year}
  {2021})}\BibitemShut {NoStop}%
\bibitem [{\citenamefont {Li}\ \emph {et~al.}(2021)\citenamefont {Li},
  \citenamefont {Huang}, \citenamefont {Li}, \citenamefont {Zhao},
  \citenamefont {Lu}, \citenamefont {Han},\ and\ \citenamefont
  {Wang}}]{li2021recent}%
  \BibitemOpen
  \bibfield  {author} {\bibinfo {author} {\bibfnamefont {W.}~\bibnamefont
  {Li}}, \bibinfo {author} {\bibfnamefont {J.}~\bibnamefont {Huang}}, \bibinfo
  {author} {\bibfnamefont {X.}~\bibnamefont {Li}}, \bibinfo {author}
  {\bibfnamefont {S.}~\bibnamefont {Zhao}}, \bibinfo {author} {\bibfnamefont
  {J.}~\bibnamefont {Lu}}, \bibinfo {author} {\bibfnamefont {Z.~V.}\
  \bibnamefont {Han}},\ and\ \bibinfo {author} {\bibfnamefont {H.}~\bibnamefont
  {Wang}},\ }\href@noop {} {\bibfield  {journal} {\bibinfo  {journal}
  {Materials Today Physics}\ }\textbf {\bibinfo {volume} {21}},\ \bibinfo
  {pages} {100504} (\bibinfo {year} {2021})}\BibitemShut {NoStop}%
\bibitem [{\citenamefont {Wickramaratne}\ \emph {et~al.}(2020)\citenamefont
  {Wickramaratne}, \citenamefont {Khmelevskyi}, \citenamefont {Agterberg},\
  and\ \citenamefont {Mazin}}]{isingprx}%
  \BibitemOpen
  \bibfield  {author} {\bibinfo {author} {\bibfnamefont {D.}~\bibnamefont
  {Wickramaratne}}, \bibinfo {author} {\bibfnamefont {S.}~\bibnamefont
  {Khmelevskyi}}, \bibinfo {author} {\bibfnamefont {D.~F.}\ \bibnamefont
  {Agterberg}},\ and\ \bibinfo {author} {\bibfnamefont {I.}~\bibnamefont
  {Mazin}},\ }\href@noop {} {\bibfield  {journal} {\bibinfo  {journal} {Phys.
  Rev. X}\ }\textbf {\bibinfo {volume} {10}},\ \bibinfo {pages} {041003}
  (\bibinfo {year} {2020})}\BibitemShut {NoStop}%
\bibitem [{\citenamefont {Divilov}\ \emph {et~al.}(2021)\citenamefont
  {Divilov}, \citenamefont {Wan}, \citenamefont {Dreher}, \citenamefont
  {B{\"o}len}, \citenamefont {S{\'a}nchez-Portal}, \citenamefont {Ugeda},\ and\
  \citenamefont {Yndur{\'a}in}}]{divilov2021magnetic}%
  \BibitemOpen
  \bibfield  {author} {\bibinfo {author} {\bibfnamefont {S.}~\bibnamefont
  {Divilov}}, \bibinfo {author} {\bibfnamefont {W.}~\bibnamefont {Wan}},
  \bibinfo {author} {\bibfnamefont {P.}~\bibnamefont {Dreher}}, \bibinfo
  {author} {\bibfnamefont {E.}~\bibnamefont {B{\"o}len}}, \bibinfo {author}
  {\bibfnamefont {D.}~\bibnamefont {S{\'a}nchez-Portal}}, \bibinfo {author}
  {\bibfnamefont {M.~M.}\ \bibnamefont {Ugeda}},\ and\ \bibinfo {author}
  {\bibfnamefont {F.}~\bibnamefont {Yndur{\'a}in}},\ }\href@noop {} {\bibfield
  {journal} {\bibinfo  {journal} {Journal of Physics: Condensed Matter}\
  }\textbf {\bibinfo {volume} {33}},\ \bibinfo {pages} {295804} (\bibinfo
  {year} {2021})}\BibitemShut {NoStop}%
\bibitem [{\citenamefont {Zhao}\ \emph {et~al.}(2019)\citenamefont {Zhao},
  \citenamefont {Lin}, \citenamefont {Xiao}, \citenamefont {Huang},
  \citenamefont {Yao}, \citenamefont {Yan}, \citenamefont {Xing}, \citenamefont
  {Zhang}, \citenamefont {Li}, \citenamefont {Hoshino} \emph
  {et~al.}}]{zhao2019disorder}%
  \BibitemOpen
  \bibfield  {author} {\bibinfo {author} {\bibfnamefont {K.}~\bibnamefont
  {Zhao}}, \bibinfo {author} {\bibfnamefont {H.}~\bibnamefont {Lin}}, \bibinfo
  {author} {\bibfnamefont {X.}~\bibnamefont {Xiao}}, \bibinfo {author}
  {\bibfnamefont {W.}~\bibnamefont {Huang}}, \bibinfo {author} {\bibfnamefont
  {W.}~\bibnamefont {Yao}}, \bibinfo {author} {\bibfnamefont {M.}~\bibnamefont
  {Yan}}, \bibinfo {author} {\bibfnamefont {Y.}~\bibnamefont {Xing}}, \bibinfo
  {author} {\bibfnamefont {Q.}~\bibnamefont {Zhang}}, \bibinfo {author}
  {\bibfnamefont {Z.-X.}\ \bibnamefont {Li}}, \bibinfo {author} {\bibfnamefont
  {S.}~\bibnamefont {Hoshino}}, \emph {et~al.},\ }\href@noop {} {\bibfield
  {journal} {\bibinfo  {journal} {Nat. Phys.}\ }\textbf {\bibinfo {volume}
  {15}},\ \bibinfo {pages} {904} (\bibinfo {year} {2019})}\BibitemShut
  {NoStop}%
\bibitem [{\citenamefont {Rubio-Verdu}\ \emph {et~al.}(2020)\citenamefont
  {Rubio-Verdu}, \citenamefont {Garcia-Garcia}, \citenamefont {Ryu},
  \citenamefont {Choi}, \citenamefont {Zaldivar}, \citenamefont {Tang},
  \citenamefont {Fan}, \citenamefont {Shen}, \citenamefont {{M}o},
  \citenamefont {Pascual},\ and\ \citenamefont
  {Ugeda}}]{rubio2020visualization}%
  \BibitemOpen
  \bibfield  {author} {\bibinfo {author} {\bibfnamefont {C.}~\bibnamefont
  {Rubio-Verdu}}, \bibinfo {author} {\bibfnamefont {A.~M.}\ \bibnamefont
  {Garcia-Garcia}}, \bibinfo {author} {\bibfnamefont {H.}~\bibnamefont {Ryu}},
  \bibinfo {author} {\bibfnamefont {D.-J.}\ \bibnamefont {Choi}}, \bibinfo
  {author} {\bibfnamefont {J.}~\bibnamefont {Zaldivar}}, \bibinfo {author}
  {\bibfnamefont {S.}~\bibnamefont {Tang}}, \bibinfo {author} {\bibfnamefont
  {B.}~\bibnamefont {Fan}}, \bibinfo {author} {\bibfnamefont {Z.-X.}\
  \bibnamefont {Shen}}, \bibinfo {author} {\bibfnamefont {S.-K.}\ \bibnamefont
  {{M}o}}, \bibinfo {author} {\bibfnamefont {J.~I.}\ \bibnamefont {Pascual}},\
  and\ \bibinfo {author} {\bibfnamefont {M.~M.}\ \bibnamefont {Ugeda}},\ }\href
  {https://doi.org/10.1021/acs.nanolett.0c01288} {\bibfield  {journal}
  {\bibinfo  {journal} {Nano Lett.}\ }\textbf {\bibinfo {volume} {20}},\
  \bibinfo {pages} {5111} (\bibinfo {year} {2020})},\ \bibinfo {note} {pMID:
  32463696},\ \Eprint
  {https://arxiv.org/abs/https://doi.org/10.1021/acs.nanolett.0c01288}
  {https://doi.org/10.1021/acs.nanolett.0c01288} \BibitemShut {NoStop}%
\bibitem [{\citenamefont {Idzuchi}\ \emph {et~al.}(2021)\citenamefont
  {Idzuchi}, \citenamefont {Pientka}, \citenamefont {Huang}, \citenamefont
  {Harada}, \citenamefont {G{\"u}l}, \citenamefont {Shin}, \citenamefont
  {Nguyen}, \citenamefont {Jo}, \citenamefont {Shindo}, \citenamefont {Cava}
  \emph {et~al.}}]{idzuchi2021unconventional}%
  \BibitemOpen
  \bibfield  {author} {\bibinfo {author} {\bibfnamefont {H.}~\bibnamefont
  {Idzuchi}}, \bibinfo {author} {\bibfnamefont {F.}~\bibnamefont {Pientka}},
  \bibinfo {author} {\bibfnamefont {K.-F.}\ \bibnamefont {Huang}}, \bibinfo
  {author} {\bibfnamefont {K.}~\bibnamefont {Harada}}, \bibinfo {author}
  {\bibfnamefont {{\"O}.}~\bibnamefont {G{\"u}l}}, \bibinfo {author}
  {\bibfnamefont {Y.~J.}\ \bibnamefont {Shin}}, \bibinfo {author}
  {\bibfnamefont {L.}~\bibnamefont {Nguyen}}, \bibinfo {author} {\bibfnamefont
  {N.}~\bibnamefont {Jo}}, \bibinfo {author} {\bibfnamefont {D.}~\bibnamefont
  {Shindo}}, \bibinfo {author} {\bibfnamefont {R.}~\bibnamefont {Cava}}, \emph
  {et~al.},\ }\href@noop {} {\bibfield  {journal} {\bibinfo  {journal} {Nat.
  Comm.}\ }\textbf {\bibinfo {volume} {12}},\ \bibinfo {pages} {1} (\bibinfo
  {year} {2021})}\BibitemShut {NoStop}%
\bibitem [{\citenamefont {Cho}\ \emph {et~al.}(2022)\citenamefont {Cho},
  \citenamefont {Lyu}, \citenamefont {An}, \citenamefont {Han}, \citenamefont
  {Lo}, \citenamefont {Ng}, \citenamefont {Hu}, \citenamefont {Gao},
  \citenamefont {Li}, \citenamefont {Huang} \emph {et~al.}}]{cho2022nodal}%
  \BibitemOpen
  \bibfield  {author} {\bibinfo {author} {\bibfnamefont {C.-w.}\ \bibnamefont
  {Cho}}, \bibinfo {author} {\bibfnamefont {J.}~\bibnamefont {Lyu}}, \bibinfo
  {author} {\bibfnamefont {L.}~\bibnamefont {An}}, \bibinfo {author}
  {\bibfnamefont {T.}~\bibnamefont {Han}}, \bibinfo {author} {\bibfnamefont
  {K.~T.}\ \bibnamefont {Lo}}, \bibinfo {author} {\bibfnamefont {C.~Y.}\
  \bibnamefont {Ng}}, \bibinfo {author} {\bibfnamefont {J.}~\bibnamefont {Hu}},
  \bibinfo {author} {\bibfnamefont {Y.}~\bibnamefont {Gao}}, \bibinfo {author}
  {\bibfnamefont {G.}~\bibnamefont {Li}}, \bibinfo {author} {\bibfnamefont
  {M.}~\bibnamefont {Huang}}, \emph {et~al.},\ }\href@noop {} {\bibfield
  {journal} {\bibinfo  {journal} {Phys. Rev. Lett.}\ }\textbf {\bibinfo
  {volume} {129}},\ \bibinfo {pages} {087002} (\bibinfo {year}
  {2022})}\BibitemShut {NoStop}%
\bibitem [{\citenamefont {Hamill}\ \emph {et~al.}(2021)\citenamefont {Hamill},
  \citenamefont {Heischmidt}, \citenamefont {Sohn}, \citenamefont {Shaffer},
  \citenamefont {Tsai}, \citenamefont {Zhang}, \citenamefont {Xi},
  \citenamefont {Suslov}, \citenamefont {Berger}, \citenamefont {Forr{\'o}}
  \emph {et~al.}}]{hamill2021two}%
  \BibitemOpen
  \bibfield  {author} {\bibinfo {author} {\bibfnamefont {A.}~\bibnamefont
  {Hamill}}, \bibinfo {author} {\bibfnamefont {B.}~\bibnamefont {Heischmidt}},
  \bibinfo {author} {\bibfnamefont {E.}~\bibnamefont {Sohn}}, \bibinfo {author}
  {\bibfnamefont {D.}~\bibnamefont {Shaffer}}, \bibinfo {author} {\bibfnamefont
  {K.-T.}\ \bibnamefont {Tsai}}, \bibinfo {author} {\bibfnamefont
  {X.}~\bibnamefont {Zhang}}, \bibinfo {author} {\bibfnamefont
  {X.}~\bibnamefont {Xi}}, \bibinfo {author} {\bibfnamefont {A.}~\bibnamefont
  {Suslov}}, \bibinfo {author} {\bibfnamefont {H.}~\bibnamefont {Berger}},
  \bibinfo {author} {\bibfnamefont {L.}~\bibnamefont {Forr{\'o}}}, \emph
  {et~al.},\ }\href@noop {} {\bibfield  {journal} {\bibinfo  {journal} {Nat.
  Phys.}\ }\textbf {\bibinfo {volume} {17}},\ \bibinfo {pages} {949} (\bibinfo
  {year} {2021})}\BibitemShut {NoStop}%
\bibitem [{\citenamefont {Das}\ and\ \citenamefont
  {Mazin}(2021)}]{das2021renormalized}%
  \BibitemOpen
  \bibfield  {author} {\bibinfo {author} {\bibfnamefont {S.}~\bibnamefont
  {Das}}\ and\ \bibinfo {author} {\bibfnamefont {I.~I.}\ \bibnamefont
  {Mazin}},\ }\href@noop {} {\bibfield  {journal} {\bibinfo  {journal} {Comput.
  Mater. Sci.}\ }\textbf {\bibinfo {volume} {200}},\ \bibinfo {pages} {110758}
  (\bibinfo {year} {2021})}\BibitemShut {NoStop}%
\bibitem [{\citenamefont {Frindt}(1972)}]{frindt1972superconductivity}%
  \BibitemOpen
  \bibfield  {author} {\bibinfo {author} {\bibfnamefont {R.}~\bibnamefont
  {Frindt}},\ }\href@noop {} {\bibfield  {journal} {\bibinfo  {journal} {Phys.
  Rev. Lett}\ }\textbf {\bibinfo {volume} {28}},\ \bibinfo {pages} {299}
  (\bibinfo {year} {1972})}\BibitemShut {NoStop}%
\bibitem [{\citenamefont {Boaknin}\ \emph {et~al.}(2003)\citenamefont
  {Boaknin}, \citenamefont {Tanatar}, \citenamefont {Paglione}, \citenamefont
  {Hawthorn}, \citenamefont {Ronning}, \citenamefont {Hill}, \citenamefont
  {Sutherland}, \citenamefont {Taillefer}, \citenamefont {Sonier},
  \citenamefont {Hayden} \emph {et~al.}}]{boaknin2003heat}%
  \BibitemOpen
  \bibfield  {author} {\bibinfo {author} {\bibfnamefont {E.}~\bibnamefont
  {Boaknin}}, \bibinfo {author} {\bibfnamefont {M.}~\bibnamefont {Tanatar}},
  \bibinfo {author} {\bibfnamefont {J.}~\bibnamefont {Paglione}}, \bibinfo
  {author} {\bibfnamefont {D.}~\bibnamefont {Hawthorn}}, \bibinfo {author}
  {\bibfnamefont {F.}~\bibnamefont {Ronning}}, \bibinfo {author} {\bibfnamefont
  {R.}~\bibnamefont {Hill}}, \bibinfo {author} {\bibfnamefont {M.}~\bibnamefont
  {Sutherland}}, \bibinfo {author} {\bibfnamefont {L.}~\bibnamefont
  {Taillefer}}, \bibinfo {author} {\bibfnamefont {J.}~\bibnamefont {Sonier}},
  \bibinfo {author} {\bibfnamefont {S.}~\bibnamefont {Hayden}}, \emph
  {et~al.},\ }\href@noop {} {\bibfield  {journal} {\bibinfo  {journal} {Phys.
  Rev. Lett.}\ }\textbf {\bibinfo {volume} {90}},\ \bibinfo {pages} {117003}
  (\bibinfo {year} {2003})}\BibitemShut {NoStop}%
\bibitem [{\citenamefont {Noat}\ \emph {et~al.}(2015)\citenamefont {Noat},
  \citenamefont {Silva-Guill{\'e}n}, \citenamefont {Cren}, \citenamefont
  {Cherkez}, \citenamefont {Brun}, \citenamefont {Pons}, \citenamefont
  {Debontridder}, \citenamefont {Roditchev}, \citenamefont {Sacks},
  \citenamefont {Cario} \emph {et~al.}}]{noat2015quasiparticle}%
  \BibitemOpen
  \bibfield  {author} {\bibinfo {author} {\bibfnamefont {Y.}~\bibnamefont
  {Noat}}, \bibinfo {author} {\bibfnamefont {J.}~\bibnamefont
  {Silva-Guill{\'e}n}}, \bibinfo {author} {\bibfnamefont {T.}~\bibnamefont
  {Cren}}, \bibinfo {author} {\bibfnamefont {V.}~\bibnamefont {Cherkez}},
  \bibinfo {author} {\bibfnamefont {C.}~\bibnamefont {Brun}}, \bibinfo {author}
  {\bibfnamefont {S.}~\bibnamefont {Pons}}, \bibinfo {author} {\bibfnamefont
  {F.}~\bibnamefont {Debontridder}}, \bibinfo {author} {\bibfnamefont
  {D.}~\bibnamefont {Roditchev}}, \bibinfo {author} {\bibfnamefont
  {W.}~\bibnamefont {Sacks}}, \bibinfo {author} {\bibfnamefont
  {L.}~\bibnamefont {Cario}}, \emph {et~al.},\ }\href@noop {} {\bibfield
  {journal} {\bibinfo  {journal} {Phys. Rev. B}\ }\textbf {\bibinfo {volume}
  {92}},\ \bibinfo {pages} {134510} (\bibinfo {year} {2015})}\BibitemShut
  {NoStop}%
\bibitem [{\citenamefont {Johannes}, \citenamefont {Mazin},\ and\ \citenamefont
  {Howells}(2006)}]{johannes2006fermi}%
  \BibitemOpen
  \bibfield  {author} {\bibinfo {author} {\bibfnamefont {M.}~\bibnamefont
  {Johannes}}, \bibinfo {author} {\bibfnamefont {I.}~\bibnamefont {Mazin}},\
  and\ \bibinfo {author} {\bibfnamefont {C.}~\bibnamefont {Howells}},\
  }\href@noop {} {\bibfield  {journal} {\bibinfo  {journal} {Phys. Rev. B}\
  }\textbf {\bibinfo {volume} {73}},\ \bibinfo {pages} {205102} (\bibinfo
  {year} {2006})}\BibitemShut {NoStop}%
\bibitem [{\citenamefont {Rossnagel}\ \emph {et~al.}(2001)\citenamefont
  {Rossnagel}, \citenamefont {Seifarth}, \citenamefont {Kipp}, \citenamefont
  {Skibowski}, \citenamefont {Vo{\ss}}, \citenamefont {Kr{\"u}ger},
  \citenamefont {Mazur},\ and\ \citenamefont {Pollmann}}]{rossnagel2001fermi}%
  \BibitemOpen
  \bibfield  {author} {\bibinfo {author} {\bibfnamefont {K.}~\bibnamefont
  {Rossnagel}}, \bibinfo {author} {\bibfnamefont {O.}~\bibnamefont {Seifarth}},
  \bibinfo {author} {\bibfnamefont {L.}~\bibnamefont {Kipp}}, \bibinfo {author}
  {\bibfnamefont {M.}~\bibnamefont {Skibowski}}, \bibinfo {author}
  {\bibfnamefont {D.}~\bibnamefont {Vo{\ss}}}, \bibinfo {author} {\bibfnamefont
  {P.}~\bibnamefont {Kr{\"u}ger}}, \bibinfo {author} {\bibfnamefont
  {A.}~\bibnamefont {Mazur}},\ and\ \bibinfo {author} {\bibfnamefont
  {J.}~\bibnamefont {Pollmann}},\ }\href@noop {} {\bibfield  {journal}
  {\bibinfo  {journal} {Phys. Rev. B}\ }\textbf {\bibinfo {volume} {64}},\
  \bibinfo {pages} {235119} (\bibinfo {year} {2001})}\BibitemShut {NoStop}%
\bibitem [{\citenamefont {Iavarone}\ \emph {et~al.}(2008)\citenamefont
  {Iavarone}, \citenamefont {Di~Capua}, \citenamefont {Karapetrov},
  \citenamefont {Koshelev}, \citenamefont {Rosenmann}, \citenamefont {Claus},
  \citenamefont {Malliakas}, \citenamefont {Kanatzidis}, \citenamefont
  {Nishizaki},\ and\ \citenamefont {Kobayashi}}]{iavarone2008effect}%
  \BibitemOpen
  \bibfield  {author} {\bibinfo {author} {\bibfnamefont {M.}~\bibnamefont
  {Iavarone}}, \bibinfo {author} {\bibfnamefont {R.}~\bibnamefont {Di~Capua}},
  \bibinfo {author} {\bibfnamefont {G.}~\bibnamefont {Karapetrov}}, \bibinfo
  {author} {\bibfnamefont {A.}~\bibnamefont {Koshelev}}, \bibinfo {author}
  {\bibfnamefont {D.}~\bibnamefont {Rosenmann}}, \bibinfo {author}
  {\bibfnamefont {H.}~\bibnamefont {Claus}}, \bibinfo {author} {\bibfnamefont
  {C.}~\bibnamefont {Malliakas}}, \bibinfo {author} {\bibfnamefont {M.~G.}\
  \bibnamefont {Kanatzidis}}, \bibinfo {author} {\bibfnamefont
  {T.}~\bibnamefont {Nishizaki}},\ and\ \bibinfo {author} {\bibfnamefont
  {N.}~\bibnamefont {Kobayashi}},\ }\href@noop {} {\bibfield  {journal}
  {\bibinfo  {journal} {Phys. Rev. B}\ }\textbf {\bibinfo {volume} {78}},\
  \bibinfo {pages} {174518} (\bibinfo {year} {2008})}\BibitemShut {NoStop}%
\bibitem [{\citenamefont {Larson}, \citenamefont {Mazin},\ and\ \citenamefont
  {Singh}(2004)}]{larson2004magnetism}%
  \BibitemOpen
  \bibfield  {author} {\bibinfo {author} {\bibfnamefont {P.}~\bibnamefont
  {Larson}}, \bibinfo {author} {\bibfnamefont {I.}~\bibnamefont {Mazin}},\ and\
  \bibinfo {author} {\bibfnamefont {D.}~\bibnamefont {Singh}},\ }\href@noop {}
  {\bibfield  {journal} {\bibinfo  {journal} {Phys. Rev. B}\ }\textbf {\bibinfo
  {volume} {69}},\ \bibinfo {pages} {064429} (\bibinfo {year}
  {2004})}\BibitemShut {NoStop}%
\bibitem [{\citenamefont {Mazin}\ \emph {et~al.}(2008)\citenamefont {Mazin},
  \citenamefont {Johannes}, \citenamefont {Boeri}, \citenamefont {Koepernik},\
  and\ \citenamefont {Singh}}]{mazin2008problems}%
  \BibitemOpen
  \bibfield  {author} {\bibinfo {author} {\bibfnamefont {I.}~\bibnamefont
  {Mazin}}, \bibinfo {author} {\bibfnamefont {M.}~\bibnamefont {Johannes}},
  \bibinfo {author} {\bibfnamefont {L.}~\bibnamefont {Boeri}}, \bibinfo
  {author} {\bibfnamefont {K.}~\bibnamefont {Koepernik}},\ and\ \bibinfo
  {author} {\bibfnamefont {D.~J.}\ \bibnamefont {Singh}},\ }\href@noop {}
  {\bibfield  {journal} {\bibinfo  {journal} {Phys. Rev. B}\ }\textbf {\bibinfo
  {volume} {78}},\ \bibinfo {pages} {085104} (\bibinfo {year}
  {2008})}\BibitemShut {NoStop}%
\bibitem [{\citenamefont {Moriya}(2012)}]{moriya2012spin}%
  \BibitemOpen
  \bibfield  {author} {\bibinfo {author} {\bibfnamefont {T.}~\bibnamefont
  {Moriya}},\ }\href@noop {} {\emph {\bibinfo {title} {Spin fluctuations in
  itinerant electron magnetism}}},\ Vol.~\bibinfo {volume} {56}\ (\bibinfo
  {publisher} {Springer Science \& Business Media},\ \bibinfo {year}
  {2012})\BibitemShut {NoStop}%
\bibitem [{\citenamefont {Das}\ \emph {et~al.}(2022)\citenamefont {Das},
  \citenamefont {Paudyal}, \citenamefont {Margine}, \citenamefont {Agterberg},\
  and\ \citenamefont {Mazin}}]{das2022electron}%
  \BibitemOpen
  \bibfield  {author} {\bibinfo {author} {\bibfnamefont {S.}~\bibnamefont
  {Das}}, \bibinfo {author} {\bibfnamefont {H.}~\bibnamefont {Paudyal}},
  \bibinfo {author} {\bibfnamefont {E.}~\bibnamefont {Margine}}, \bibinfo
  {author} {\bibfnamefont {D.}~\bibnamefont {Agterberg}},\ and\ \bibinfo
  {author} {\bibfnamefont {I.}~\bibnamefont {Mazin}},\ }\href@noop {}
  {\bibfield  {journal} {\bibinfo  {journal} {arXiv preprint arXiv:2210.00745}\
  } (\bibinfo {year} {2022})}\BibitemShut {NoStop}%
\bibitem [{\citenamefont {Costa}, \citenamefont {Costa},\ and\ \citenamefont
  {Fern{\'a}ndez-Rossier}(2022)}]{costa2022ising}%
  \BibitemOpen
  \bibfield  {author} {\bibinfo {author} {\bibfnamefont {A.~T.}\ \bibnamefont
  {Costa}}, \bibinfo {author} {\bibfnamefont {M.}~\bibnamefont {Costa}},\ and\
  \bibinfo {author} {\bibfnamefont {J.}~\bibnamefont {Fern{\'a}ndez-Rossier}},\
  }\href@noop {} {\bibfield  {journal} {\bibinfo  {journal} {Phys. Rev. B}\
  }\textbf {\bibinfo {volume} {105}},\ \bibinfo {pages} {224412} (\bibinfo
  {year} {2022})}\BibitemShut {NoStop}%
\bibitem [{\citenamefont {Zheng}\ and\ \citenamefont
  {Feng}(2019)}]{zheng2019electron}%
  \BibitemOpen
  \bibfield  {author} {\bibinfo {author} {\bibfnamefont {F.}~\bibnamefont
  {Zheng}}\ and\ \bibinfo {author} {\bibfnamefont {J.}~\bibnamefont {Feng}},\
  }\href@noop {} {\bibfield  {journal} {\bibinfo  {journal} {Phys. Rev. B}\
  }\textbf {\bibinfo {volume} {99}},\ \bibinfo {pages} {161119} (\bibinfo
  {year} {2019})}\BibitemShut {NoStop}%
\bibitem [{\citenamefont {Wan}\ \emph {et~al.}(2022{\natexlab{a}})\citenamefont
  {Wan}, \citenamefont {Dreher}, \citenamefont {Mu{\~n}oz-Segovia},
  \citenamefont {Harsh}, \citenamefont {Guo}, \citenamefont
  {Mart{\'\i}nez-Galera}, \citenamefont {Guinea}, \citenamefont {de~Juan},\
  and\ \citenamefont {Ugeda}}]{wan2022observation}%
  \BibitemOpen
  \bibfield  {author} {\bibinfo {author} {\bibfnamefont {W.}~\bibnamefont
  {Wan}}, \bibinfo {author} {\bibfnamefont {P.}~\bibnamefont {Dreher}},
  \bibinfo {author} {\bibfnamefont {D.}~\bibnamefont {Mu{\~n}oz-Segovia}},
  \bibinfo {author} {\bibfnamefont {R.}~\bibnamefont {Harsh}}, \bibinfo
  {author} {\bibfnamefont {H.}~\bibnamefont {Guo}}, \bibinfo {author}
  {\bibfnamefont {A.~J.}\ \bibnamefont {Mart{\'\i}nez-Galera}}, \bibinfo
  {author} {\bibfnamefont {F.}~\bibnamefont {Guinea}}, \bibinfo {author}
  {\bibfnamefont {F.}~\bibnamefont {de~Juan}},\ and\ \bibinfo {author}
  {\bibfnamefont {M.~M.}\ \bibnamefont {Ugeda}},\ }\href@noop {} {\bibfield
  {journal} {\bibinfo  {journal} {Adv. Mater.}\ ,\ \bibinfo {pages} {2206078}}
  (\bibinfo {year} {2022}{\natexlab{a}})}\BibitemShut {NoStop}%
\bibitem [{\citenamefont {Leroux}\ \emph {et~al.}(2015)\citenamefont {Leroux},
  \citenamefont {Errea}, \citenamefont {Le~Tacon}, \citenamefont {Souliou},
  \citenamefont {Garbarino}, \citenamefont {Cario}, \citenamefont {Bosak},
  \citenamefont {Mauri}, \citenamefont {Calandra},\ and\ \citenamefont
  {Rodi{\`e}re}}]{leroux2015strong}%
  \BibitemOpen
  \bibfield  {author} {\bibinfo {author} {\bibfnamefont {M.}~\bibnamefont
  {Leroux}}, \bibinfo {author} {\bibfnamefont {I.}~\bibnamefont {Errea}},
  \bibinfo {author} {\bibfnamefont {M.}~\bibnamefont {Le~Tacon}}, \bibinfo
  {author} {\bibfnamefont {S.-M.}\ \bibnamefont {Souliou}}, \bibinfo {author}
  {\bibfnamefont {G.}~\bibnamefont {Garbarino}}, \bibinfo {author}
  {\bibfnamefont {L.}~\bibnamefont {Cario}}, \bibinfo {author} {\bibfnamefont
  {A.}~\bibnamefont {Bosak}}, \bibinfo {author} {\bibfnamefont
  {F.}~\bibnamefont {Mauri}}, \bibinfo {author} {\bibfnamefont
  {M.}~\bibnamefont {Calandra}},\ and\ \bibinfo {author} {\bibfnamefont
  {P.}~\bibnamefont {Rodi{\`e}re}},\ }\href@noop {} {\bibfield  {journal}
  {\bibinfo  {journal} {Phys. Rev. B}\ }\textbf {\bibinfo {volume} {92}},\
  \bibinfo {pages} {140303} (\bibinfo {year} {2015})}\BibitemShut {NoStop}%
\bibitem [{\citenamefont {Cho}\ \emph {et~al.}(2018)\citenamefont {Cho},
  \citenamefont {Ko{\'n}czykowski}, \citenamefont {Teknowijoyo}, \citenamefont
  {Tanatar}, \citenamefont {Guss}, \citenamefont {Gartin}, \citenamefont
  {Wilde}, \citenamefont {Kreyssig}, \citenamefont {McQueeney}, \citenamefont
  {Goldman} \emph {et~al.}}]{cho2018using}%
  \BibitemOpen
  \bibfield  {author} {\bibinfo {author} {\bibfnamefont {K.}~\bibnamefont
  {Cho}}, \bibinfo {author} {\bibfnamefont {M.}~\bibnamefont
  {Ko{\'n}czykowski}}, \bibinfo {author} {\bibfnamefont {S.}~\bibnamefont
  {Teknowijoyo}}, \bibinfo {author} {\bibfnamefont {M.~A.}\ \bibnamefont
  {Tanatar}}, \bibinfo {author} {\bibfnamefont {J.}~\bibnamefont {Guss}},
  \bibinfo {author} {\bibfnamefont {P.}~\bibnamefont {Gartin}}, \bibinfo
  {author} {\bibfnamefont {J.~M.}\ \bibnamefont {Wilde}}, \bibinfo {author}
  {\bibfnamefont {A.}~\bibnamefont {Kreyssig}}, \bibinfo {author}
  {\bibfnamefont {R.}~\bibnamefont {McQueeney}}, \bibinfo {author}
  {\bibfnamefont {A.~I.}\ \bibnamefont {Goldman}}, \emph {et~al.},\ }\href@noop
  {} {\bibfield  {journal} {\bibinfo  {journal} {Nat. Comm.}\ }\textbf
  {\bibinfo {volume} {9}},\ \bibinfo {pages} {1} (\bibinfo {year}
  {2018})}\BibitemShut {NoStop}%
\bibitem [{\citenamefont {Kuzmanovi{\'c}}\ \emph {et~al.}(2022)\citenamefont
  {Kuzmanovi{\'c}}, \citenamefont {Dvir}, \citenamefont {Leboeuf},
  \citenamefont {Ili{\'c}}, \citenamefont {Haim}, \citenamefont {M{\"o}ckli},
  \citenamefont {Kramer}, \citenamefont {Khodas}, \citenamefont {Houzet},
  \citenamefont {Meyer} \emph {et~al.}}]{kuzmanovic2022tunneling}%
  \BibitemOpen
  \bibfield  {author} {\bibinfo {author} {\bibfnamefont {M.}~\bibnamefont
  {Kuzmanovi{\'c}}}, \bibinfo {author} {\bibfnamefont {T.}~\bibnamefont
  {Dvir}}, \bibinfo {author} {\bibfnamefont {D.}~\bibnamefont {Leboeuf}},
  \bibinfo {author} {\bibfnamefont {S.}~\bibnamefont {Ili{\'c}}}, \bibinfo
  {author} {\bibfnamefont {M.}~\bibnamefont {Haim}}, \bibinfo {author}
  {\bibfnamefont {D.}~\bibnamefont {M{\"o}ckli}}, \bibinfo {author}
  {\bibfnamefont {S.}~\bibnamefont {Kramer}}, \bibinfo {author} {\bibfnamefont
  {M.}~\bibnamefont {Khodas}}, \bibinfo {author} {\bibfnamefont
  {M.}~\bibnamefont {Houzet}}, \bibinfo {author} {\bibfnamefont
  {J.}~\bibnamefont {Meyer}}, \emph {et~al.},\ }\href@noop {} {\bibfield
  {journal} {\bibinfo  {journal} {Phys. Rev. B}\ }\textbf {\bibinfo {volume}
  {106}},\ \bibinfo {pages} {184514} (\bibinfo {year} {2022})}\BibitemShut
  {NoStop}%
\bibitem [{\citenamefont {Wan}\ \emph {et~al.}(2022{\natexlab{b}})\citenamefont
  {Wan}, \citenamefont {Wickramaratne}, \citenamefont {Dreher}, \citenamefont
  {Harsh}, \citenamefont {Mazin},\ and\ \citenamefont
  {Ugeda}}]{wan2022nontrivial}%
  \BibitemOpen
  \bibfield  {author} {\bibinfo {author} {\bibfnamefont {W.}~\bibnamefont
  {Wan}}, \bibinfo {author} {\bibfnamefont {D.}~\bibnamefont {Wickramaratne}},
  \bibinfo {author} {\bibfnamefont {P.}~\bibnamefont {Dreher}}, \bibinfo
  {author} {\bibfnamefont {R.}~\bibnamefont {Harsh}}, \bibinfo {author}
  {\bibfnamefont {I.~I.}\ \bibnamefont {Mazin}},\ and\ \bibinfo {author}
  {\bibfnamefont {M.~M.}\ \bibnamefont {Ugeda}},\ }\href@noop {} {\bibfield
  {journal} {\bibinfo  {journal} {Adv. Mater.}\ }\textbf {\bibinfo {volume}
  {34}},\ \bibinfo {pages} {2200492} (\bibinfo {year}
  {2022}{\natexlab{b}})}\BibitemShut {NoStop}%
\bibitem [{\citenamefont {Wickramaratne}\ and\ \citenamefont
  {Mazin}(2022)}]{nbse2_fractal_natcomm}%
  \BibitemOpen
  \bibfield  {author} {\bibinfo {author} {\bibfnamefont {D.}~\bibnamefont
  {Wickramaratne}}\ and\ \bibinfo {author} {\bibfnamefont {I.}~\bibnamefont
  {Mazin}},\ }\href@noop {} {\bibfield  {journal} {\bibinfo  {journal} {Nat.
  Comm.}\ }\textbf {\bibinfo {volume} {13}},\ \bibinfo {pages} {1} (\bibinfo
  {year} {2022})}\BibitemShut {NoStop}%
\bibitem [{\citenamefont {Wickramaratne}\ \emph {et~al.}(2021)\citenamefont
  {Wickramaratne}, \citenamefont {Haim}, \citenamefont {Khodas},\ and\
  \citenamefont {Mazin}}]{nbse2_proximity}%
  \BibitemOpen
  \bibfield  {author} {\bibinfo {author} {\bibfnamefont {D.}~\bibnamefont
  {Wickramaratne}}, \bibinfo {author} {\bibfnamefont {M.}~\bibnamefont {Haim}},
  \bibinfo {author} {\bibfnamefont {M.}~\bibnamefont {Khodas}},\ and\ \bibinfo
  {author} {\bibfnamefont {I.~I.}\ \bibnamefont {Mazin}},\ }\href
  {https://doi.org/10.1103/PhysRevB.104.L060501} {\bibfield  {journal}
  {\bibinfo  {journal} {Phys. Rev. B}\ }\textbf {\bibinfo {volume} {104}},\
  \bibinfo {pages} {L060501} (\bibinfo {year} {2021})}\BibitemShut {NoStop}%
\bibitem [{\citenamefont {M{\"o}ckli}\ and\ \citenamefont
  {Khodas}(2020)}]{mockli2020Ising}%
  \BibitemOpen
  \bibfield  {author} {\bibinfo {author} {\bibfnamefont {D.}~\bibnamefont
  {M{\"o}ckli}}\ and\ \bibinfo {author} {\bibfnamefont {M.}~\bibnamefont
  {Khodas}},\ }\href@noop {} {\bibfield  {journal} {\bibinfo  {journal} {Phys.
  Rev. B}\ }\textbf {\bibinfo {volume} {101}},\ \bibinfo {pages} {014510}
  (\bibinfo {year} {2020})}\BibitemShut {NoStop}%
\bibitem [{\citenamefont {Kang}\ \emph {et~al.}(2022)\citenamefont {Kang},
  \citenamefont {Berger}, \citenamefont {Watanabe}, \citenamefont {Taniguchi},
  \citenamefont {Forró}, \citenamefont {Shan},\ and\ \citenamefont
  {Mak}}]{kang2022van}%
  \BibitemOpen
  \bibfield  {author} {\bibinfo {author} {\bibfnamefont {K.}~\bibnamefont
  {Kang}}, \bibinfo {author} {\bibfnamefont {H.}~\bibnamefont {Berger}},
  \bibinfo {author} {\bibfnamefont {K.}~\bibnamefont {Watanabe}}, \bibinfo
  {author} {\bibfnamefont {T.}~\bibnamefont {Taniguchi}}, \bibinfo {author}
  {\bibfnamefont {L.}~\bibnamefont {Forró}}, \bibinfo {author} {\bibfnamefont
  {J.}~\bibnamefont {Shan}},\ and\ \bibinfo {author} {\bibfnamefont {K.~F.}\
  \bibnamefont {Mak}},\ }\href {https://doi.org/10.1021/acs.nanolett.2c01640}
  {\bibfield  {journal} {\bibinfo  {journal} {Nano Lett.}\ }\textbf {\bibinfo
  {volume} {22}},\ \bibinfo {pages} {5510} (\bibinfo {year} {2022})},\ \bibinfo
  {note} {pMID: 35736540},\ \Eprint
  {https://arxiv.org/abs/https://doi.org/10.1021/acs.nanolett.2c01640}
  {https://doi.org/10.1021/acs.nanolett.2c01640} \BibitemShut {NoStop}%
\bibitem [{\citenamefont {Ai}\ \emph {et~al.}(2021)\citenamefont {Ai},
  \citenamefont {Zhang}, \citenamefont {Yang}, \citenamefont {Xie},
  \citenamefont {Yang}, \citenamefont {Jia}, \citenamefont {Zhang},
  \citenamefont {Liu}, \citenamefont {Li}, \citenamefont {Leng} \emph
  {et~al.}}]{ai2021van}%
  \BibitemOpen
  \bibfield  {author} {\bibinfo {author} {\bibfnamefont {L.}~\bibnamefont
  {Ai}}, \bibinfo {author} {\bibfnamefont {E.}~\bibnamefont {Zhang}}, \bibinfo
  {author} {\bibfnamefont {J.}~\bibnamefont {Yang}}, \bibinfo {author}
  {\bibfnamefont {X.}~\bibnamefont {Xie}}, \bibinfo {author} {\bibfnamefont
  {Y.}~\bibnamefont {Yang}}, \bibinfo {author} {\bibfnamefont {Z.}~\bibnamefont
  {Jia}}, \bibinfo {author} {\bibfnamefont {Y.}~\bibnamefont {Zhang}}, \bibinfo
  {author} {\bibfnamefont {S.}~\bibnamefont {Liu}}, \bibinfo {author}
  {\bibfnamefont {Z.}~\bibnamefont {Li}}, \bibinfo {author} {\bibfnamefont
  {P.}~\bibnamefont {Leng}}, \emph {et~al.},\ }\href@noop {} {\bibfield
  {journal} {\bibinfo  {journal} {Nat. Comm.}\ }\textbf {\bibinfo {volume}
  {12}},\ \bibinfo {pages} {1} (\bibinfo {year} {2021})}\BibitemShut {NoStop}%
\bibitem [{\citenamefont {Kang}\ \emph {et~al.}(2021)\citenamefont {Kang},
  \citenamefont {Jiang}, \citenamefont {Berger}, \citenamefont {Watanabe},
  \citenamefont {Taniguchi}, \citenamefont {Forr{\'o}}, \citenamefont {Shan},\
  and\ \citenamefont {Mak}}]{kang2021giant}%
  \BibitemOpen
  \bibfield  {author} {\bibinfo {author} {\bibfnamefont {K.}~\bibnamefont
  {Kang}}, \bibinfo {author} {\bibfnamefont {S.}~\bibnamefont {Jiang}},
  \bibinfo {author} {\bibfnamefont {H.}~\bibnamefont {Berger}}, \bibinfo
  {author} {\bibfnamefont {K.}~\bibnamefont {Watanabe}}, \bibinfo {author}
  {\bibfnamefont {T.}~\bibnamefont {Taniguchi}}, \bibinfo {author}
  {\bibfnamefont {L.}~\bibnamefont {Forr{\'o}}}, \bibinfo {author}
  {\bibfnamefont {J.}~\bibnamefont {Shan}},\ and\ \bibinfo {author}
  {\bibfnamefont {K.~F.}\ \bibnamefont {Mak}},\ }\href@noop {} {\bibfield
  {journal} {\bibinfo  {journal} {arXiv preprint arXiv:2101.01327}\ } (\bibinfo
  {year} {2021})}\BibitemShut {NoStop}%
\bibitem [{\citenamefont {Wu}\ \emph {et~al.}(2022)\citenamefont {Wu},
  \citenamefont {Xu}, \citenamefont {Haley}, \citenamefont {Weber},
  \citenamefont {Acharya}, \citenamefont {Maniv}, \citenamefont {Qiu},
  \citenamefont {Aczel}, \citenamefont {Settineri}, \citenamefont {Neaton}
  \emph {et~al.}}]{wu2022highly}%
  \BibitemOpen
  \bibfield  {author} {\bibinfo {author} {\bibfnamefont {S.}~\bibnamefont
  {Wu}}, \bibinfo {author} {\bibfnamefont {Z.}~\bibnamefont {Xu}}, \bibinfo
  {author} {\bibfnamefont {S.~C.}\ \bibnamefont {Haley}}, \bibinfo {author}
  {\bibfnamefont {S.~F.}\ \bibnamefont {Weber}}, \bibinfo {author}
  {\bibfnamefont {A.}~\bibnamefont {Acharya}}, \bibinfo {author} {\bibfnamefont
  {E.}~\bibnamefont {Maniv}}, \bibinfo {author} {\bibfnamefont
  {Y.}~\bibnamefont {Qiu}}, \bibinfo {author} {\bibfnamefont {A.}~\bibnamefont
  {Aczel}}, \bibinfo {author} {\bibfnamefont {N.~S.}\ \bibnamefont
  {Settineri}}, \bibinfo {author} {\bibfnamefont {J.~B.}\ \bibnamefont
  {Neaton}}, \emph {et~al.},\ }\href@noop {} {\bibfield  {journal} {\bibinfo
  {journal} {Phys. Rev. X}\ }\textbf {\bibinfo {volume} {12}},\ \bibinfo
  {pages} {021003} (\bibinfo {year} {2022})}\BibitemShut {NoStop}%
\bibitem [{\citenamefont {Haley}\ \emph {et~al.}(2021)\citenamefont {Haley},
  \citenamefont {Maniv}, \citenamefont {Cookmeyer}, \citenamefont
  {Torres-Londono}, \citenamefont {Aravinth}, \citenamefont {Moore},\ and\
  \citenamefont {Analytis}}]{haley2021long}%
  \BibitemOpen
  \bibfield  {author} {\bibinfo {author} {\bibfnamefont {S.~C.}\ \bibnamefont
  {Haley}}, \bibinfo {author} {\bibfnamefont {E.}~\bibnamefont {Maniv}},
  \bibinfo {author} {\bibfnamefont {T.}~\bibnamefont {Cookmeyer}}, \bibinfo
  {author} {\bibfnamefont {S.}~\bibnamefont {Torres-Londono}}, \bibinfo
  {author} {\bibfnamefont {M.}~\bibnamefont {Aravinth}}, \bibinfo {author}
  {\bibfnamefont {J.}~\bibnamefont {Moore}},\ and\ \bibinfo {author}
  {\bibfnamefont {J.~G.}\ \bibnamefont {Analytis}},\ }\href@noop {} {\bibfield
  {journal} {\bibinfo  {journal} {arXiv preprint arXiv:2111.09882}\ } (\bibinfo
  {year} {2021})}\BibitemShut {NoStop}%
\bibitem [{\citenamefont {Xie}\ \emph {et~al.}(2022)\citenamefont {Xie},
  \citenamefont {Husremovic}, \citenamefont {Gonzalez}, \citenamefont {Craig},\
  and\ \citenamefont {Bediako}}]{xie2022structure}%
  \BibitemOpen
  \bibfield  {author} {\bibinfo {author} {\bibfnamefont {L.~S.}\ \bibnamefont
  {Xie}}, \bibinfo {author} {\bibfnamefont {S.}~\bibnamefont {Husremovic}},
  \bibinfo {author} {\bibfnamefont {O.}~\bibnamefont {Gonzalez}}, \bibinfo
  {author} {\bibfnamefont {I.~M.}\ \bibnamefont {Craig}},\ and\ \bibinfo
  {author} {\bibfnamefont {D.~K.}\ \bibnamefont {Bediako}},\ }\href@noop {}
  {\bibfield  {journal} {\bibinfo  {journal} {J. Am. Chem. Soc.}\ } (\bibinfo
  {year} {2022})}\BibitemShut {NoStop}%
\bibitem [{\citenamefont {Kashihara}, \citenamefont {Nishida},\ and\
  \citenamefont {Yoshioka}(1979)}]{kashihara1979upper}%
  \BibitemOpen
  \bibfield  {author} {\bibinfo {author} {\bibfnamefont {Y.}~\bibnamefont
  {Kashihara}}, \bibinfo {author} {\bibfnamefont {A.}~\bibnamefont {Nishida}},\
  and\ \bibinfo {author} {\bibfnamefont {H.}~\bibnamefont {Yoshioka}},\
  }\href@noop {} {\bibfield  {journal} {\bibinfo  {journal} {J. Phys. Soc.
  Japan}\ }\textbf {\bibinfo {volume} {46}},\ \bibinfo {pages} {1112} (\bibinfo
  {year} {1979})}\BibitemShut {NoStop}%
\bibitem [{\citenamefont {Coleman}\ \emph {et~al.}(1983)\citenamefont
  {Coleman}, \citenamefont {Eiserman}, \citenamefont {Hillenius}, \citenamefont
  {Mitchell},\ and\ \citenamefont {Vicent}}]{coleman1983dimensional}%
  \BibitemOpen
  \bibfield  {author} {\bibinfo {author} {\bibfnamefont {R.}~\bibnamefont
  {Coleman}}, \bibinfo {author} {\bibfnamefont {G.}~\bibnamefont {Eiserman}},
  \bibinfo {author} {\bibfnamefont {S.}~\bibnamefont {Hillenius}}, \bibinfo
  {author} {\bibfnamefont {A.}~\bibnamefont {Mitchell}},\ and\ \bibinfo
  {author} {\bibfnamefont {J.}~\bibnamefont {Vicent}},\ }\href@noop {}
  {\bibfield  {journal} {\bibinfo  {journal} {Phys. Rev. B}\ }\textbf {\bibinfo
  {volume} {27}},\ \bibinfo {pages} {125} (\bibinfo {year} {1983})}\BibitemShut
  {NoStop}%
\bibitem [{\citenamefont {Samuely}\ \emph {et~al.}(2021)\citenamefont
  {Samuely}, \citenamefont {Szab{\'o}}, \citenamefont
  {Ka{\v{c}}mar{\v{c}}{\'\i}k}, \citenamefont {Meerschaut}, \citenamefont
  {Cario}, \citenamefont {Jansen}, \citenamefont {Cren}, \citenamefont
  {Kuzmiak}, \citenamefont {{\v{S}}ofranko},\ and\ \citenamefont
  {Samuely}}]{samuely2021extreme}%
  \BibitemOpen
  \bibfield  {author} {\bibinfo {author} {\bibfnamefont {P.}~\bibnamefont
  {Samuely}}, \bibinfo {author} {\bibfnamefont {P.}~\bibnamefont {Szab{\'o}}},
  \bibinfo {author} {\bibfnamefont {J.}~\bibnamefont
  {Ka{\v{c}}mar{\v{c}}{\'\i}k}}, \bibinfo {author} {\bibfnamefont
  {A.}~\bibnamefont {Meerschaut}}, \bibinfo {author} {\bibfnamefont
  {L.}~\bibnamefont {Cario}}, \bibinfo {author} {\bibfnamefont
  {A.}~\bibnamefont {Jansen}}, \bibinfo {author} {\bibfnamefont
  {T.}~\bibnamefont {Cren}}, \bibinfo {author} {\bibfnamefont {M.}~\bibnamefont
  {Kuzmiak}}, \bibinfo {author} {\bibfnamefont {O.}~\bibnamefont
  {{\v{S}}ofranko}},\ and\ \bibinfo {author} {\bibfnamefont {T.}~\bibnamefont
  {Samuely}},\ }\href@noop {} {\bibfield  {journal} {\bibinfo  {journal} {Phys.
  Rev. B}\ }\textbf {\bibinfo {volume} {104}},\ \bibinfo {pages} {224507}
  (\bibinfo {year} {2021})}\BibitemShut {NoStop}%
\bibitem [{\citenamefont {Leriche}\ \emph {et~al.}(2021)\citenamefont
  {Leriche}, \citenamefont {Palacio-Morales}, \citenamefont {Campetella},
  \citenamefont {Tresca}, \citenamefont {Sasaki}, \citenamefont {Brun},
  \citenamefont {Debontridder}, \citenamefont {David}, \citenamefont {Arfaoui},
  \citenamefont {{\v{S}}ofranko} \emph {et~al.}}]{leriche2021misfit}%
  \BibitemOpen
  \bibfield  {author} {\bibinfo {author} {\bibfnamefont {R.~T.}\ \bibnamefont
  {Leriche}}, \bibinfo {author} {\bibfnamefont {A.}~\bibnamefont
  {Palacio-Morales}}, \bibinfo {author} {\bibfnamefont {M.}~\bibnamefont
  {Campetella}}, \bibinfo {author} {\bibfnamefont {C.}~\bibnamefont {Tresca}},
  \bibinfo {author} {\bibfnamefont {S.}~\bibnamefont {Sasaki}}, \bibinfo
  {author} {\bibfnamefont {C.}~\bibnamefont {Brun}}, \bibinfo {author}
  {\bibfnamefont {F.}~\bibnamefont {Debontridder}}, \bibinfo {author}
  {\bibfnamefont {P.}~\bibnamefont {David}}, \bibinfo {author} {\bibfnamefont
  {I.}~\bibnamefont {Arfaoui}}, \bibinfo {author} {\bibfnamefont
  {O.}~\bibnamefont {{\v{S}}ofranko}}, \emph {et~al.},\ }\href@noop {}
  {\bibfield  {journal} {\bibinfo  {journal} {Adv. Func. Mater.}\ }\textbf
  {\bibinfo {volume} {31}},\ \bibinfo {pages} {2007706} (\bibinfo {year}
  {2021})}\BibitemShut {NoStop}%
\bibitem [{\citenamefont {Devarakonda}\ \emph {et~al.}(2021)\citenamefont
  {Devarakonda}, \citenamefont {Suzuki}, \citenamefont {Fang}, \citenamefont
  {Zhu}, \citenamefont {Graf}, \citenamefont {Kriener}, \citenamefont {Fu},
  \citenamefont {Kaxiras},\ and\ \citenamefont
  {Checkelsky}}]{devarakonda2021signatures}%
  \BibitemOpen
  \bibfield  {author} {\bibinfo {author} {\bibfnamefont {A.}~\bibnamefont
  {Devarakonda}}, \bibinfo {author} {\bibfnamefont {T.}~\bibnamefont {Suzuki}},
  \bibinfo {author} {\bibfnamefont {S.}~\bibnamefont {Fang}}, \bibinfo {author}
  {\bibfnamefont {J.}~\bibnamefont {Zhu}}, \bibinfo {author} {\bibfnamefont
  {D.}~\bibnamefont {Graf}}, \bibinfo {author} {\bibfnamefont {M.}~\bibnamefont
  {Kriener}}, \bibinfo {author} {\bibfnamefont {L.}~\bibnamefont {Fu}},
  \bibinfo {author} {\bibfnamefont {E.}~\bibnamefont {Kaxiras}},\ and\ \bibinfo
  {author} {\bibfnamefont {J.}~\bibnamefont {Checkelsky}},\ }\href@noop {}
  {\bibfield  {journal} {\bibinfo  {journal} {Nature}\ }\textbf {\bibinfo
  {volume} {599}},\ \bibinfo {pages} {51} (\bibinfo {year} {2021})}\BibitemShut
  {NoStop}%
\bibitem [{\citenamefont {Ma}\ \emph {et~al.}(2018)\citenamefont {Ma},
  \citenamefont {Pan}, \citenamefont {Guo}, \citenamefont {Zhang},
  \citenamefont {Wang}, \citenamefont {Hu}, \citenamefont {Mu}, \citenamefont
  {Huang},\ and\ \citenamefont {Xie}}]{ma2018unusual}%
  \BibitemOpen
  \bibfield  {author} {\bibinfo {author} {\bibfnamefont {Y.}~\bibnamefont
  {Ma}}, \bibinfo {author} {\bibfnamefont {J.}~\bibnamefont {Pan}}, \bibinfo
  {author} {\bibfnamefont {C.}~\bibnamefont {Guo}}, \bibinfo {author}
  {\bibfnamefont {X.}~\bibnamefont {Zhang}}, \bibinfo {author} {\bibfnamefont
  {L.}~\bibnamefont {Wang}}, \bibinfo {author} {\bibfnamefont {T.}~\bibnamefont
  {Hu}}, \bibinfo {author} {\bibfnamefont {G.}~\bibnamefont {Mu}}, \bibinfo
  {author} {\bibfnamefont {F.}~\bibnamefont {Huang}},\ and\ \bibinfo {author}
  {\bibfnamefont {X.}~\bibnamefont {Xie}},\ }\href@noop {} {\bibfield
  {journal} {\bibinfo  {journal} {npj Quantum Mater.}\ }\textbf {\bibinfo
  {volume} {3}},\ \bibinfo {pages} {1} (\bibinfo {year} {2018})}\BibitemShut
  {NoStop}%
\bibitem [{\citenamefont {Prober}, \citenamefont {Schwall},\ and\ \citenamefont
  {Beasley}(1980)}]{prober1980upper}%
  \BibitemOpen
  \bibfield  {author} {\bibinfo {author} {\bibfnamefont {D.}~\bibnamefont
  {Prober}}, \bibinfo {author} {\bibfnamefont {R.}~\bibnamefont {Schwall}},\
  and\ \bibinfo {author} {\bibfnamefont {M.}~\bibnamefont {Beasley}},\
  }\href@noop {} {\bibfield  {journal} {\bibinfo  {journal} {Phys. Rev. B}\
  }\textbf {\bibinfo {volume} {21}},\ \bibinfo {pages} {2717} (\bibinfo {year}
  {1980})}\BibitemShut {NoStop}%
\bibitem [{\citenamefont {Gamble}\ \emph {et~al.}(1970)\citenamefont {Gamble},
  \citenamefont {DiSalvo}, \citenamefont {Klemm},\ and\ \citenamefont
  {Geballe}}]{gamble1970superconductivity}%
  \BibitemOpen
  \bibfield  {author} {\bibinfo {author} {\bibfnamefont {F.}~\bibnamefont
  {Gamble}}, \bibinfo {author} {\bibfnamefont {F.}~\bibnamefont {DiSalvo}},
  \bibinfo {author} {\bibfnamefont {R.}~\bibnamefont {Klemm}},\ and\ \bibinfo
  {author} {\bibfnamefont {T.}~\bibnamefont {Geballe}},\ }\href@noop {}
  {\bibfield  {journal} {\bibinfo  {journal} {Science}\ }\textbf {\bibinfo
  {volume} {168}},\ \bibinfo {pages} {568} (\bibinfo {year}
  {1970})}\BibitemShut {NoStop}%
\bibitem [{\citenamefont {Xing}\ \emph {et~al.}(2021)\citenamefont {Xing},
  \citenamefont {Yang}, \citenamefont {Ge}, \citenamefont {Yan}, \citenamefont
  {Luo}, \citenamefont {Ji}, \citenamefont {Yang}, \citenamefont {Li},
  \citenamefont {Wang}, \citenamefont {Liu} \emph
  {et~al.}}]{xing2021extrinsic}%
  \BibitemOpen
  \bibfield  {author} {\bibinfo {author} {\bibfnamefont {Y.}~\bibnamefont
  {Xing}}, \bibinfo {author} {\bibfnamefont {P.}~\bibnamefont {Yang}}, \bibinfo
  {author} {\bibfnamefont {J.}~\bibnamefont {Ge}}, \bibinfo {author}
  {\bibfnamefont {J.}~\bibnamefont {Yan}}, \bibinfo {author} {\bibfnamefont
  {J.}~\bibnamefont {Luo}}, \bibinfo {author} {\bibfnamefont {H.}~\bibnamefont
  {Ji}}, \bibinfo {author} {\bibfnamefont {Z.}~\bibnamefont {Yang}}, \bibinfo
  {author} {\bibfnamefont {Y.}~\bibnamefont {Li}}, \bibinfo {author}
  {\bibfnamefont {Z.}~\bibnamefont {Wang}}, \bibinfo {author} {\bibfnamefont
  {Y.}~\bibnamefont {Liu}}, \emph {et~al.},\ }\href@noop {} {\bibfield
  {journal} {\bibinfo  {journal} {Nano Lett.}\ }\textbf {\bibinfo {volume}
  {21}},\ \bibinfo {pages} {7486} (\bibinfo {year} {2021})}\BibitemShut
  {NoStop}%
\bibitem [{\citenamefont {Ribak}\ \emph {et~al.}(2020)\citenamefont {Ribak},
  \citenamefont {Skiff}, \citenamefont {{M}ograbi}, \citenamefont {Rout},
  \citenamefont {Fischer}, \citenamefont {Ruhman}, \citenamefont {Chashka},
  \citenamefont {Dagan},\ and\ \citenamefont {Kanigel}}]{ribak2020chiral}%
  \BibitemOpen
  \bibfield  {author} {\bibinfo {author} {\bibfnamefont {A.}~\bibnamefont
  {Ribak}}, \bibinfo {author} {\bibfnamefont {R.~M.}\ \bibnamefont {Skiff}},
  \bibinfo {author} {\bibfnamefont {M.}~\bibnamefont {{M}ograbi}}, \bibinfo
  {author} {\bibfnamefont {P.}~\bibnamefont {Rout}}, \bibinfo {author}
  {\bibfnamefont {M.}~\bibnamefont {Fischer}}, \bibinfo {author} {\bibfnamefont
  {J.}~\bibnamefont {Ruhman}}, \bibinfo {author} {\bibfnamefont
  {K.}~\bibnamefont {Chashka}}, \bibinfo {author} {\bibfnamefont
  {Y.}~\bibnamefont {Dagan}},\ and\ \bibinfo {author} {\bibfnamefont
  {A.}~\bibnamefont {Kanigel}},\ }\href@noop {} {\bibfield  {journal} {\bibinfo
   {journal} {Sci. Adv.}\ }\textbf {\bibinfo {volume} {6}},\ \bibinfo {pages}
  {eaax9480} (\bibinfo {year} {2020})}\BibitemShut {NoStop}%
\bibitem [{\citenamefont {Achari}\ \emph {et~al.}(2022)\citenamefont {Achari},
  \citenamefont {Bekaert}, \citenamefont {Sreepal}, \citenamefont {Orekhov},
  \citenamefont {Kumaravadivel}, \citenamefont {Kim}, \citenamefont
  {Gauquelin}, \citenamefont {Balakrishna~Pillai}, \citenamefont {Verbeeck},
  \citenamefont {Peeters} \emph {et~al.}}]{achari2022alternating}%
  \BibitemOpen
  \bibfield  {author} {\bibinfo {author} {\bibfnamefont {A.}~\bibnamefont
  {Achari}}, \bibinfo {author} {\bibfnamefont {J.}~\bibnamefont {Bekaert}},
  \bibinfo {author} {\bibfnamefont {V.}~\bibnamefont {Sreepal}}, \bibinfo
  {author} {\bibfnamefont {A.}~\bibnamefont {Orekhov}}, \bibinfo {author}
  {\bibfnamefont {P.}~\bibnamefont {Kumaravadivel}}, \bibinfo {author}
  {\bibfnamefont {M.}~\bibnamefont {Kim}}, \bibinfo {author} {\bibfnamefont
  {N.}~\bibnamefont {Gauquelin}}, \bibinfo {author} {\bibfnamefont
  {P.}~\bibnamefont {Balakrishna~Pillai}}, \bibinfo {author} {\bibfnamefont
  {J.}~\bibnamefont {Verbeeck}}, \bibinfo {author} {\bibfnamefont {F.~M.}\
  \bibnamefont {Peeters}}, \emph {et~al.},\ }\href@noop {} {\bibfield
  {journal} {\bibinfo  {journal} {Nano Lett.}\ }\textbf {\bibinfo {volume}
  {22}},\ \bibinfo {pages} {6268} (\bibinfo {year} {2022})}\BibitemShut
  {NoStop}%
\bibitem [{\citenamefont {Tanaka}\ \emph {et~al.}(2020)\citenamefont {Tanaka},
  \citenamefont {Matsuoka}, \citenamefont {Nakano}, \citenamefont {Wang},
  \citenamefont {Sasakura}, \citenamefont {Kobayashi},\ and\ \citenamefont
  {Iwasa}}]{tanaka2020superconducting}%
  \BibitemOpen
  \bibfield  {author} {\bibinfo {author} {\bibfnamefont {Y.}~\bibnamefont
  {Tanaka}}, \bibinfo {author} {\bibfnamefont {H.}~\bibnamefont {Matsuoka}},
  \bibinfo {author} {\bibfnamefont {M.}~\bibnamefont {Nakano}}, \bibinfo
  {author} {\bibfnamefont {Y.}~\bibnamefont {Wang}}, \bibinfo {author}
  {\bibfnamefont {S.}~\bibnamefont {Sasakura}}, \bibinfo {author}
  {\bibfnamefont {K.}~\bibnamefont {Kobayashi}},\ and\ \bibinfo {author}
  {\bibfnamefont {Y.}~\bibnamefont {Iwasa}},\ }\href@noop {} {\bibfield
  {journal} {\bibinfo  {journal} {Nano Lett.}\ }\textbf {\bibinfo {volume}
  {20}},\ \bibinfo {pages} {1725} (\bibinfo {year} {2020})}\BibitemShut
  {NoStop}%
\bibitem [{\citenamefont {Samuely}\ \emph {et~al.}(2023)\citenamefont
  {Samuely}, \citenamefont {Wickramaratne}, \citenamefont {Gmitra},
  \citenamefont {Jaouen}, \citenamefont {Šofranko}, \citenamefont {Volavka},
  \citenamefont {Kuzmiak}, \citenamefont {Haniš}, \citenamefont {Szabó},
  \citenamefont {Monney}, \citenamefont {Kremer}, \citenamefont {Fèvre},
  \citenamefont {Bertran}, \citenamefont {Cren}, \citenamefont {Sasaki},
  \citenamefont {Cario}, \citenamefont {Calandra}, \citenamefont {Mazin},\ and\
  \citenamefont {Samuely}}]{samuely2023protection}%
  \BibitemOpen
  \bibfield  {author} {\bibinfo {author} {\bibfnamefont {T.}~\bibnamefont
  {Samuely}}, \bibinfo {author} {\bibfnamefont {D.}~\bibnamefont
  {Wickramaratne}}, \bibinfo {author} {\bibfnamefont {M.}~\bibnamefont
  {Gmitra}}, \bibinfo {author} {\bibfnamefont {T.}~\bibnamefont {Jaouen}},
  \bibinfo {author} {\bibfnamefont {O.}~\bibnamefont {Šofranko}}, \bibinfo
  {author} {\bibfnamefont {D.}~\bibnamefont {Volavka}}, \bibinfo {author}
  {\bibfnamefont {M.}~\bibnamefont {Kuzmiak}}, \bibinfo {author} {\bibfnamefont
  {J.}~\bibnamefont {Haniš}}, \bibinfo {author} {\bibfnamefont
  {P.}~\bibnamefont {Szabó}}, \bibinfo {author} {\bibfnamefont
  {C.}~\bibnamefont {Monney}}, \bibinfo {author} {\bibfnamefont
  {G.}~\bibnamefont {Kremer}}, \bibinfo {author} {\bibfnamefont {P.~L.}\
  \bibnamefont {Fèvre}}, \bibinfo {author} {\bibfnamefont {F.}~\bibnamefont
  {Bertran}}, \bibinfo {author} {\bibfnamefont {T.}~\bibnamefont {Cren}},
  \bibinfo {author} {\bibfnamefont {S.}~\bibnamefont {Sasaki}}, \bibinfo
  {author} {\bibfnamefont {L.}~\bibnamefont {Cario}}, \bibinfo {author}
  {\bibfnamefont {M.}~\bibnamefont {Calandra}}, \bibinfo {author}
  {\bibfnamefont {I.~I.}\ \bibnamefont {Mazin}},\ and\ \bibinfo {author}
  {\bibfnamefont {P.}~\bibnamefont {Samuely}},\ }\href@noop {} {\  (\bibinfo
  {year} {2023})},\ \Eprint {https://arxiv.org/abs/2304.03074}
  {arXiv:2304.03074 [cond-mat.supr-con]} \BibitemShut {NoStop}%
\bibitem [{\citenamefont {Ng}\ and\ \citenamefont
  {McQueen}(2022)}]{ng2022misfit}%
  \BibitemOpen
  \bibfield  {author} {\bibinfo {author} {\bibfnamefont {N.}~\bibnamefont
  {Ng}}\ and\ \bibinfo {author} {\bibfnamefont {T.~M.}\ \bibnamefont
  {McQueen}},\ }\href@noop {} {\bibfield  {journal} {\bibinfo  {journal} {APL
  Mater.}\ }\textbf {\bibinfo {volume} {10}},\ \bibinfo {pages} {100901}
  (\bibinfo {year} {2022})}\BibitemShut {NoStop}%
\bibitem [{\citenamefont {Klemm}(2015)}]{klemm2015pristine}%
  \BibitemOpen
  \bibfield  {author} {\bibinfo {author} {\bibfnamefont {R.~A.}\ \bibnamefont
  {Klemm}},\ }\href@noop {} {\bibfield  {journal} {\bibinfo  {journal} {Phys.
  C: Supercond. Appl.}\ }\textbf {\bibinfo {volume} {514}},\ \bibinfo {pages}
  {86} (\bibinfo {year} {2015})}\BibitemShut {NoStop}%
\bibitem [{\citenamefont {Brown}\ and\ \citenamefont
  {Beerntsen}(1965)}]{brown1965layer}%
  \BibitemOpen
  \bibfield  {author} {\bibinfo {author} {\bibfnamefont {B.~E.}\ \bibnamefont
  {Brown}}\ and\ \bibinfo {author} {\bibfnamefont {D.~J.}\ \bibnamefont
  {Beerntsen}},\ }\href@noop {} {\bibfield  {journal} {\bibinfo  {journal}
  {Acta Crystallogr.}\ }\textbf {\bibinfo {volume} {18}},\ \bibinfo {pages}
  {31} (\bibinfo {year} {1965})}\BibitemShut {NoStop}%
\end{thebibliography}

%
\end{document}